\documentclass[preprint,12pt]{elsarticle}

\usepackage[T1]{fontenc}
\usepackage[english]{babel}
\usepackage{graphicx}
\usepackage{subcaption}
\usepackage{amsmath}
\usepackage{amssymb}
\usepackage{booktabs}
\usepackage{array}
\usepackage{bicaption}
\usepackage{float}
\usepackage{url}
\usepackage{xcolor}

\usepackage{algorithm}
\usepackage[noend]{algpseudocode}
\usepackage{placeins}%
\newcommand{\bx}{\mathbf x}
\newcommand{\bv}{\mathbf v}
\newcommand{\bF}{\mathbf F}
\newcommand{\bb}{\hat{\mathbf b}}
\newcommand{\Hel}{\mathcal H}
\newcommand{\dbar}{\bar d}

\newlength{\PaperFigurePanelWidth}
\newlength{\PaperFigureStackWidth}
\newlength{\PaperFigureFullWidth}
\newlength{\PaperFigureWideWidth}
\newlength{\PaperFigureTwoThirdsWidth}
\newcommand{\PaperFigureFull}[1]{\includegraphics[width=\PaperFigureFullWidth]{#1}}
\newcommand{\PaperFigureWide}[1]{\includegraphics[width=\PaperFigureWideWidth]{#1}}
\newcommand{\PaperFigureInPanel}[1]{\includegraphics[width=\linewidth]{#1}}
\newcommand{\PaperSchematic}[1]{\includegraphics[trim=2bp 2bp 2bp 2bp]{#1}}

\journal{Computer Physics Communications}

\begin{document}

\begin{frontmatter}

\title{Direct, Collision-Operator-Free Simulation of Electron--Ion Transport in Magnetized and Unmagnetized Plasmas}

\author[aff1]{Keheng Zhu}
\author[aff1]{Jian Liu\corref{cor1}}
\ead{liu_jian@sdu.edu.cn}
\author[aff1]{Chaozhou Mou}
\author[aff1]{Senran Lin}
\author[aff2]{Wei Zhang}
\cortext[cor1]{Corresponding author.}
\address[aff1]{Weihai Institute for Interdisciplinary Research, Shandong University, Weihai 264209, China}
\address[aff2]{FusionSage Group, Shanghai 200000, China}

\begin{abstract}
Accurate electron--ion collision models are central to predicting transport in fusion and space plasmas, yet collision operators and Monte Carlo collision models assume that a charged particle meets one partner at a time and that its successive encounters are independent. However, an electron that turns back along the magnetic field passes the same ions again. We solve the Newton--Lorentz equations for test electrons among static ions with a prescribed Debye-screened potential. The method combines force summation within an interaction radius, a time-symmetric particle pusher, and adaptive time stepping, enabling stable and scalable simulations in unmagnetized and magnetized plasmas. Using simulation-based measures of momentum relaxation and cross-field diffusion, we compare the simulations with a reference in the same model that treats every encounter as independent. Without a magnetic field the scattering rate reproduces the binary-collision rate for the same potential to within $2\%$. When the gyroradius is smaller than the Debye length, the scattering rate stays within $10\%$ of the reference, but the cross-field diffusion coefficient exceeds it by a factor that grows with the field, up to 17. The excess comes from repeated passes of the same ions, whose guiding-center displacements add up: when the ions are replaced by new random ones many times per scattering time, so that a returning electron meets different ions, most of the excess disappears. The framework therefore provides a computational benchmark for testing and improving reduced collision operators and transport models.
\end{abstract}


\end{frontmatter}

\section{Introduction}
Electron--ion collisions convert microscopic plasma dynamics into macroscopic transport. They set momentum-relaxation rates, effective diffusion coefficients, and the closure parameters used in reduced plasma models. These quantities matter in magnetic-confinement, laboratory, and space plasmas, where weakly collisional magnetized transport controls long-time particle evolution. Reduced models do not follow the forces between the particles. They replace these forces by a collision operator, which rests on two assumptions: an encounter involves one partner at a time, and successive encounters are independent. A Coulomb logarithm summarizes the range of impact parameters that contribute. The practical question is therefore whether reduced transport coefficients remain quantitatively reliable when the underlying interaction is the long-range Coulomb force. Transport coefficients computed from particle orbits make neither assumption and so test both: in a specified microscopic field, all nearby particles act at once and a particle can meet the same partners again.

Without a magnetic field, both assumptions are well justified. Debye shielding limits the effective range of Coulomb fields in a plasma \cite{huckel1923theorie}. The kinetic evolution is described by the Boltzmann equation \cite{boltzmann1872weitere}; for long-range interactions dominated by small-angle scattering, its collision term is commonly approximated in Fokker--Planck form \cite{fokkerMittlereEnergieRotierender1914,planck1917satz}. The corresponding Coulomb theory can be expressed through Rosenbluth potentials \cite{rosenbluthFokkerPlanckEquationInverseSquare1957} and, for binary collisions with impact-parameter cutoffs, through the Landau operator \cite{landau1965collected}. The Balescu--Lenard formulation incorporates collective effects in a kinetic description \cite{lenard1960bogoliubov,balescu1960irreversible}. Escande, Elskens and Doveil showed that a Debye-dressed pair interaction gives a transport expression that converges over all impact parameters and reduces to Rutherford scattering at small impact parameter \cite{escandeUniformDerivationCoulomb2015}. Because the individual deflections are weak, the transport effect of the simultaneous interactions of a particle with many others is well approximated by a sum of independent binary estimates \cite{escandeUniformDerivationCoulomb2015}. In the test-electron limit with a stationary ion background, these descriptions reduce to a Lorentz-like model \cite{dongCollisionTermUniformly2023}, in which the Coulomb logarithm represents the separation between short- and long-distance scales; this is the setting of the present work.

In a magnetic field, ions that pass the electron farther away than its gyroradius also move it across the field, but the theories that include them still treat each encounter as independent. Classical magnetized transport describes cross-field diffusion as a collision-driven random walk with a step of one gyroradius $\rho$ per collision time, giving $D_\perp \sim \nu\rho^2\propto B^{-2}$, where $\nu$ is the collision frequency and $B$ the magnetic-field strength. Each step is a close encounter, one that turns the velocity. When the gyroradius is smaller than the Debye length $\lambda_D$, ions farther away than the gyroradius also act on the electron. A distant encounter, one at an impact parameter between $\rho$ and $\lambda_D$, hardly turns the velocity. The Coulomb logarithm of velocity scattering is therefore cut off at $\rho$ instead of $\lambda_D$ \cite{montgomeryMagneticFieldDependence1974}. A distant encounter still moves the guiding center, through the $\mathbf E\times\mathbf B$ drift around the ion. In plasmas of a single species these distant encounters dominate cross-field transport \cite{oneilNewTheoryTransport1985}, and the measured test-particle diffusion exceeds the classical value by about a factor of ten \cite{andereggTestParticleTransport1997}. Like the classical theory, the theory of these distant encounters counts each encounter once, as independent of the others \cite{oneilNewTheoryTransport1985}. The history of plasma transport also illustrates the boundaries of the classical picture: early experiments on magnetized arc discharges motivated Bohm diffusion with a weaker $B^{-1}$ scaling \cite{bohm1949characteristics}, whereas magnetic geometry and trapped-particle effects led to neoclassical transport theory \cite{hinton1985neoclassical,hintonTheoryPlasmaTransport1976}. Magnetic fields can also alter the elementary scattering dynamics themselves \cite{jiangTransverseRutherfordScattering2022}.

Most computational collision models are built on these reduced descriptions. In gyrokinetic simulation, a practical collision operator must conserve particle number, momentum and energy and satisfy an H theorem \cite{sugamaModernGyrokineticFormulation2017}. Donnel et al.\ implemented and verified a linearized multi-species operator with these properties in the full-F code GYSELA \cite{donnelMultispeciesCollisionalOperator2019}. Binary-collision Monte Carlo models \cite{takizukaBinaryCollisionModel1977}, cumulative small-angle schemes of Nanbu type \cite{nanbuTheoryCumulativeSmallangle1997,nanbuWeightedParticlesCoulomb1998}, and approaches based on stochastic differential equations \cite{cadjanLangevinApproachPlasma1999,wuWeaklyConvergentStochastic2023a} are designed to reproduce Lorentz, Landau, or Fokker--Planck dynamics efficiently, and their numerical accuracy has been studied in detail \cite{dimitsUnderstandingAccuracyNanbu2009a}. They are valuable tools, but they are built to reproduce reduced collision operators, and all of them treat the successive encounters of a particle as independent events.

However, even with distant encounters included, independent encounters can underestimate cross-field transport. In a single-species plasma, Dubin showed that repeated collisions of the same pair of particles multiply the test-particle diffusion due to distant encounters by three \cite{dubinTestParticleDiffusion1997,dubinCollisionalTransportNonneutral1998}. Theories that follow each particle along its unperturbed helix leave these collisions out. The reason is that in a magnetic field successive encounters need not be independent. Along the field an electron moves almost freely; when its parallel velocity reverses, it returns along nearly the same field line and passes the same ions again. When the gyroradius is below the Debye length, the returns can add up. Polyakov argued that returns to the same static screened center increase cross-field diffusion once the gyroradius is smaller than the screening length \cite{polyakovChangeDueScatteringact1986}. Memory effects of this kind also change magnetotransport in two-dimensional Lorentz gases \cite{bobylevTwoDimensionalMagnetotransport1995,dmitrievClassicalMechanismNegative2001} and in smooth static potentials \cite{songLinearMagnetoresistanceMetals2015}. In molecular dynamics of magnetized one-component plasmas, $D_\perp$ falls more slowly than $B^{-2}$ once the gyroradius is below the Debye length \cite{baalrudTransportRegimesSpanning2017,valvoNumericalStudyTransverse2022} and follows a Bohm-like $B^{-1}$ at strong fields \cite{ottDiffusionStronglyCoupled2011}. Two-dimensional simulations of electrons and ions keep all particles in the plane perpendicular to the magnetic field, so no electron moves along it. When the electrons collide only with ions, their diffusion across the field follows a binary-collision estimate \cite{nishimuraNbodyPlasmaSimulation2025}.

No one has measured how much the returns change electron--ion transport in a plasma. For electrons among static ions with a gyroradius below the Debye length, it is not known whether independent encounters still give the scattering rate and the cross-field diffusion coefficient, or how large their error is. Baalrud and Daligault proposed repeated collisions as the cause of the slow decrease of $D_\perp$ with $B$ without measuring their effect \cite{baalrudTransportRegimesSpanning2017}. A generalized Boltzmann theory for strongly magnetized plasmas leaves repeated collisions out \cite{joseGeneralizedBoltzmannKinetic2020}, and no rigorous collision operator is available when the same pair of particles collides several times \cite{kennedyCoulombCollisionsStrongly2021}. A measurement faces three obstacles. (i) Independent encounters must be evaluated in the same model as the orbits: the analytic formulas use a different potential, truncation and orbits, and these differences would enter the comparison. (ii) The orbits must contain the same encounters as the reference, which samples impact parameters down to zero. The orbits must therefore be followed through encounters at every impact parameter, down to head-on passages, where the force diverges. (iii) The ions must stay in place along the whole path of a returning electron, more than $10^4$ Debye lengths along the field in a run.

In this work we compute electron--ion transport coefficients from test-electron orbits in a prescribed background of static, Debye-screened ions, and address each obstacle. The force on each test electron is computed at every time step as the direct vector sum of the Debye-screened fields from ions within an interaction radius $R_c$ around it. The orbit equations are integrated without a small-deflection expansion, and no collision operator or sequence of scattering events is prescribed. (i) Alongside these transport runs, a second calculation in the same model gives the independent-encounter reference: passes of an electron by a single ion, integrated with the same step, potential, interaction radius and field, are summed as independent events. (ii) The test-electron trajectories are integrated with a time-symmetric particle pusher whose step is set by a time transformation and shrinks in proportion to the distance to the nearest ion; it follows Coulomb orbits down to head-on passages (Sections~\ref{subsec:numerical-method} and~\ref{subsec:rutherford}). This approach follows the broader geometric-integration strategy used for charged-particle dynamics \cite{zhangVolumepreservingAlgorithmSecular2015a,heVolumepreservingAlgorithmsCharged2015} and in structure-preserving particle-in-cell methods for kinetic plasma simulation \cite{xiaoStructurePreserving2018,xiaoExplicitStructurePreserving2021}. (iii) The ions are generated cell by cell from a seed, so the background is fixed and unbounded without being stored, and a returning electron meets the same ions.

Transport coefficients are then obtained from trajectory statistics. Estimating diffusion from the mean-squared displacement, running diffusion coefficients, or Lagrangian velocity correlations of test-particle trajectories is well established in studies of transport in stochastic and turbulent fields \cite{vladDiffusionIntrinsicTrapping1998,naulinDispersionIdealParticles1999,milovanovStochasticDynamicsFractional2001}. We use two ensemble-averaged metrics to connect the trajectories to transport theory: the pitch-angle scattering rate, which is an effective electron--ion collision frequency, and the perpendicular diffusion coefficient of the guiding center in the magnetized case.

Where independent encounters should hold, the calculation reproduces their predictions: without a magnetic field the scattering rate agrees with the binary-collision rate computed for the same truncated potential, and in the magnetized runs it stays within $10\%$ of the independent-encounter reference. The cross-field diffusion coefficient differs: independent encounters underestimate it by a factor that grows with the field, up to 17 at a gyroradius of $0.011\lambda_D$. Two tests trace the excess to repeated passes. Splitting each trajectory where its velocity along the field reverses shows that the guiding-center displacements on successive passes over the same stretch of field line point the same way (Fig.~\ref{fig:excursions}a). Renewing the ion configuration every few hundredths of a scattering time limits the passes of the same ions to about two and removes most of the excess (Fig.~\ref{fig:excursions}b). The results provide benchmark values for collision models of electron--ion transport in plasmas whose ion configuration persists while an electron returns along the field.

Section~\ref{sec:model} describes the model, the numerical method and a single-ion verification, and Section~\ref{subsec:transport-benchmarks} the observables, the calculations compared and the splitting of trajectories. Section~\ref{sec:results} presents the transport results, Section~\ref{sec:discussion} explains why the excess grows with the field and relates the results to earlier work and to collision operators, and Section~\ref{sec:conclusion} summarizes the work.

\section{Simulation Framework for Collisional Transport}
\label{sec:model}
Classical collision models are useful because they reduce long-range Coulomb dynamics to tractable operators, but that same reduction treats successive encounters as independent. We therefore reverse the usual logic: instead of starting from a collision operator and inferring transport, we resolve the screened force of the surrounding ions directly and let the transport coefficient emerge from the trajectories. Section~\ref{subsec:model} defines the model, Section~\ref{subsec:numerical-method} the numerical method, and Section~\ref{subsec:rutherford} a verification against Rutherford scattering. Section~\ref{subsec:transport-benchmarks} turns the trajectories into observables that can be compared with transport theory. Table~\ref{tab:symbols} lists the symbols used in more than one section.

\begin{table}[!htbp]
    \centering
    \caption{Symbols used in more than one section, with the place where each is defined.}
    \label{tab:symbols}
    \footnotesize
    \setlength{\tabcolsep}{5pt}
    \begin{tabular}{@{}lll@{}}
      \toprule
      Symbol & Meaning & Defined in \\
      \midrule
      $n$, $T_e$ & density, electron temperature & Section~\ref{subsec:model} \\
      $\lambda_D$, $N_D$ & Debye length, ions per Debye sphere & Section~\ref{subsec:model} \\
      $R_c=c\lambda_D$, $N_c$ & interaction radius, mean number of ions in it & Eq.~\eqref{eq:superposed-field} \\
      $\dbar=n^{-1/3}$ & mean ion spacing & Section~\ref{subsec:numerical-method} \\
      $v_0$ & initial speed of the electrons & Section~\ref{subsec:model} \\
      $b_0$, $\kappa$ & $90^\circ$ impact parameter, $\kappa=m_ev_0^2b_0$ & Eq.~\eqref{eq:b0} \\
      $b$ & impact parameter & Section~\ref{subsec:rutherford} \\
      $\boldsymbol B$, $\bb$ & magnetic field, its direction & Section~\ref{subsec:model} \\
      $\Omega$, $\tau_g$ & cyclotron frequency $eB/m_e$, gyroperiod & Section~\ref{subsec:model} \\
      $\rho$, $\rho_v$ & gyroradius $v_\perp/\Omega$, and $v_0/\Omega$ & Section~\ref{subsec:model} \\
      \addlinespace
      $\Delta\tilde t$ & step in the transformed time (unit $\kappa/v_0$) & Section~\ref{subsubsec:adaptive-time-stepping} \\
      $g$, $W$ & step-setting function, variable tracking it & Eq.~\eqref{ttl:eq:g}, Alg.~\ref{ttl:alg} \\
      $\Delta t_{\max}$, $g_\infty$ & largest physical step, $g_\infty=\Delta\tilde t/\Delta t_{\max}$ & Eq.~\eqref{ttl:eq:g} \\
      $\ell$ & margin of the neighbor list & Alg.~\ref{ttl:alg} \\
      \addlinespace
      $\mu$ & pitch-angle cosine & Section~\ref{subsec:obs-nu} \\
      $C(\tau)$, $\nu$ & direction autocorrelation, pitch-angle scattering rate & Eq.~\eqref{eq:corr} \\
      $\nu_0$, $\ln\Lambda_{\rm eff}$ & binary-collision rate, its Coulomb logarithm & Eq.~\eqref{eq:lnL-eff} \\
      $\mathbf X$, $D_\perp$ & guiding center, cross-field diffusion coefficient & Eqs.~\eqref{eq:gc}--\eqref{eq:Dperp} \\
      $D_0$ & unit of diffusion, $\nu_0v_0^2/(3\Omega^2)$ & Section~\ref{subsec:obs-D} \\
      $\nu^{\rm IE}$, $D_\perp^{\rm IE}$ & independent-encounter reference & Section~\ref{subsec:ie-reference} \\
      $\nu_1$, $t_{\rm end}$ & rate that sets the run lengths, run length & Table~\ref{tab:runs} \\
      \addlinespace
      $\Delta\mathbf G_{i,k}$ & displacement of $\mathbf X$ of electron $i$ between reversals of $v_\parallel$ & Eq.~\eqref{eq:excursions} \\
      $\tau_c$ & renewal interval of the ion configuration & Section~\ref{subsec:ie-reference} \\
      \bottomrule
    \end{tabular}
\end{table}

\subsection{Direct Debye-Screened Test-Electron Dynamics}
\label{subsec:model}

Our physical model is designed to study collisional transport by following the electron motion under the direct action of many surrounding ions (simultaneous interactions), rather than reducing the interaction to a sequence of independent binary collisions. We therefore consider a spatially homogeneous, singly ionized hydrogen plasma ($Z=1$), and track the trajectories of test electrons moving through a fixed ion background. The ion charge is $q_i=Ze$ and the electron charge is $q_e=-e$, where $e$ is the elementary charge. The electron and ion number densities are $n_e$ and $n_i$; for $Z=1$, quasineutrality gives $n_e=n_i\equiv n$. The electron temperature is $T_e$. Because the ion mass is much larger than the electron mass ($m_i \gg m_e$), the ion motion is neglected during the electron trajectory integration and the static-ion approximation ($v_i=0$) is adopted.

To represent the collective screening effect of the background electrons on each ion, we describe each ion by the Debye-screened Coulomb potential \cite{huckel1923theorie}. In this model, the surrounding electrons redistribute in response to the ion's charge, reducing the effective range of the ion's electric field on the scale of the Debye length $\lambda_D$. The corresponding electric potential and electric field are written as
\begin{equation}
\phi(r)=\frac{q_i}{4\pi\epsilon_{0}r}e^{-r/\lambda_{D}}, \quad
\mathbf{E}(r)=\frac{q_i}{4\pi\epsilon_{0}}
\left(\frac{1}{r^{2}}+\frac{1}{\lambda_{D}r}\right)
e^{-r/\lambda_{D}}\hat{\mathbf{r}},
\label{eq:debye-electric}
\end{equation}
where $r$ is the distance from the ion, $\hat{\mathbf{r}}$ is the unit vector pointing from the ion to the field point, and
\begin{equation}
\lambda_D = \sqrt{\frac{\epsilon_0 k_\textrm{B}T_e}{n_e q_e^2}}
\end{equation}
is the Debye length, where $\epsilon_0$ is the vacuum permittivity and $k_\textrm{B}$ the Boltzmann constant. The potential is an effective pairwise interaction \cite{escandeUniformDerivationCoulomb2015}: its screening is prescribed and static and is not computed self-consistently.

At each time step, the instantaneous electric field acting on test electron $j$ is obtained by summing the contributions from all ions located within a sphere of radius $R_c=c\,\lambda_D$ centered on that electron:
\begin{equation}
\boldsymbol{E}_j(\boldsymbol{r}_j)=
\sum_{i:\,r_{ji} \leq R_c}
\frac{q_i}{4 \pi \epsilon_{0}}
\frac{1}{r_{ji}^{2}}
\left(\frac{1}{r_{ji}}+\frac{1}{\lambda_{D}}\right)
\exp\left(-\frac{r_{ji}}{\lambda_{D}}\right)
\boldsymbol{r}_{ji},
\label{eq:superposed-field}
\end{equation}
where $\boldsymbol{r}_{ji}=\boldsymbol{r}_{j}-\boldsymbol{r}_{i}$ is the relative position vector from ion $i$ to the test electron and $r_{ji}=|\boldsymbol{r}_{ji}|$. Because $q_e<0$, the force $q_e\boldsymbol{E}_j$ attracts the electron toward each ion. We call $R_c$ the interaction radius. The ions form a fixed random arrangement with mean density $n$ (Algorithm~\ref{ttl:alg}), and the sphere holds on average
\begin{equation}
N_c=\frac{4}{3}\pi n R_c^3
\end{equation}
ions; Section~\ref{sec:results} gives the interaction radii of the runs. For $c=1$, $N_c$ is the Debye number $N_D=\frac{4}{3}\pi n\lambda_D^3$, which satisfies $N_D\gg1$ for all parameters considered here.

The electron acceleration is then determined from the Lorentz force in a uniform external magnetic field $\boldsymbol{B}=B\hat{\boldsymbol{z}}$,
\begin{equation}
\boldsymbol{a}_j(\boldsymbol{r}_j,\boldsymbol{v}_j,t)
=
\frac{q_e}{m_e}
\left(
\boldsymbol{E}_j+\boldsymbol{v}_j\times\boldsymbol{B}
\right).
\label{eq:electron-acceleration}
\end{equation}
For $B\neq0$ the electron gyrates with the cyclotron frequency $\Omega=eB/m_e$ and the gyroperiod $\tau_g=2\pi/\Omega$, and its gyroradius is $\rho=v_\perp/\Omega$, where $v_\perp$ is the speed perpendicular to $\boldsymbol{B}$. The test electrons start with speed $v_0$ in an isotropic random direction. In the transport runs $v_0$ equals the thermal speed $v_{th}=(3k_\textrm{B}T_e/m_e)^{1/2}$, and we label the field strengths by the gyroradius $\rho_v=v_0/\Omega$ of an electron that moves perpendicular to $\boldsymbol B$. The impact parameter for $90^\circ$ deflection of an electron with speed $v_0$ by a single ion is
\begin{equation}
b_0=\frac{|q_i q_e|}{4\pi\epsilon_0 m_e v_0^{2}}.
\label{eq:b0}
\end{equation}

\begin{figure}[!htbp]
    \centering
    \PaperSchematic{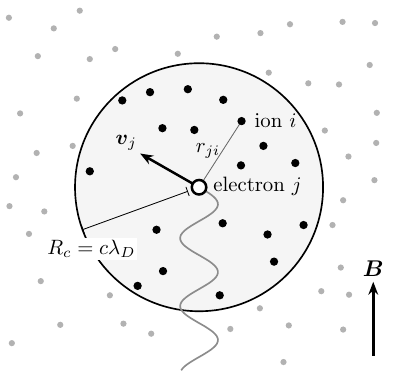}
    \caption{Physical setting of the simulation, drawn in a plane that contains $\boldsymbol B$ and not to scale. The incident electron interacts with surrounding static ions within a sphere of radius $R_c=c\lambda_D$ through Debye-screened Coulomb forces. The sphere is centered on the electron and moves with it. Filled dots are the ions within the sphere, which make up the field of Eq.~\eqref{eq:superposed-field}; gray dots are ions outside it. The arrow is the velocity $\boldsymbol v_j$, whose initial direction is isotropic, and the gray line is the earlier path, which gyrates about $\boldsymbol B$ in the magnetized runs.}
    \label{fig:physical-setting}
\end{figure}

With this formulation, the transport properties are not imposed through a prescribed collision operator. Instead, they emerge directly from the simulated trajectories under the combined action of the Debye-screened Coulomb forces of all ions in the sphere and the external magnetic field. The field is evaluated at every time step along the actual orbit, and Eq.~\eqref{eq:electron-acceleration} is integrated without a small-deflection expansion. Figure~\ref{fig:physical-setting} illustrates the physical setting.

The static-ion approximation, the prescribed static screening, the truncation of the field at $R_c$, and the omitted electron--electron collisions and electromagnetic fluctuations define the principal limitations of the model. The magnetized results therefore apply where the ion configuration persists while an electron returns along the field, as for static scatterers \cite{polyakovChangeDueScatteringact1986} and approximately for plasmas with heavy, cold ions.

\begin{figure}[!htbp]
    \centering
    \PaperFigureFull{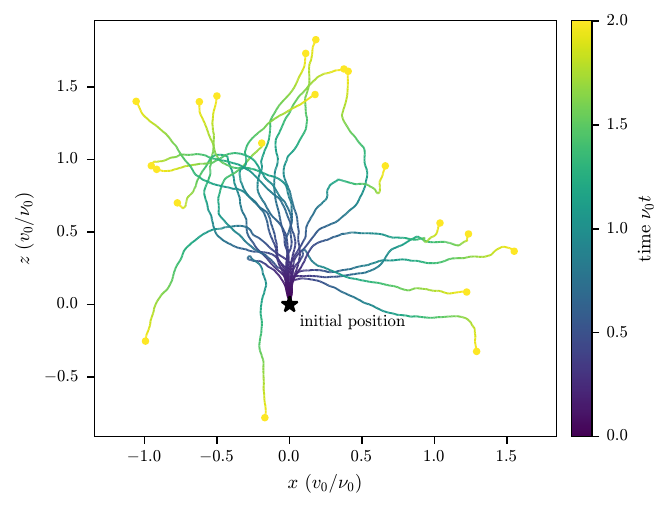}
    \caption{Representative test-electron trajectories in the unmagnetized case. Twenty electrons are followed for two collision times; each path is shifted to a common starting point and rotated so that its initial velocity points along $z$. The paths are projected on a plane that contains the initial velocity and are colored by the time $t$ in units of $1/\nu_0$, where $\nu_0$ is the binary-collision rate of Section~\ref{subsec:obs-nu}; lengths are in units of the distance $v_0/\nu_0$ traveled in one collision time. The density and temperature are those of Section~\ref{sec:results}, and $R_c=3\lambda_D$. The trajectories are mostly smooth and continuously bent by long-range Debye-screened Coulomb forces. A few close encounters produce sharper directional changes.}
    \label{fig:unmagnetized-trajectories}
\end{figure}

Figure~\ref{fig:unmagnetized-trajectories} provides a representative visualization of particle motion in the unmagnetized case. The trajectories are typically smooth and continuously curved over long distances, reflecting the cumulative action of long-range Coulomb forces from many surrounding ions. At the same time, a small number of close approaches produce sharper changes in direction, so rare near-field interactions can still generate large velocity deflections. Resolving both contributions numerically requires an integration strategy that remains stable over long trajectories while adapting to widely separated interaction scales.

\subsection{Numerical Strategy: Stable Long-Time Integration Across Scales}
\label{subsec:numerical-method}\label{ttl:method}

The direct formulation places three related requirements on the numerical method. First, the particle advance must remain reliable over the long trajectories required to measure transport coefficients. Second, the integration must resolve both cumulative weak forcing and rare close encounters without using an unnecessarily small time step throughout the simulation. Third, the force evaluation must remain computationally tractable for large trajectory ensembles. We address these requirements using, respectively, a time-symmetric particle integrator together with midpoint-consistent field evaluation, adaptive time stepping, and a moving interaction-sphere implementation parallelized over the electron ensemble.

To maintain long-time fidelity, we integrate the Newton--Lorentz equations~(\ref{eq:electron-acceleration}) by splitting each step into simpler motions, like the relativistic volume-preserving algorithm (RVPA), which preserves phase-space volume and is therefore well suited to long-time charged-particle integration \cite{zhangVolumepreservingAlgorithmSecular2015a,heVolumepreservingAlgorithmsCharged2015}. Here each step, as in RVPA, consists of a half drift, a kick at the midpoint of the step, and a second half drift (Section~\ref{subsubsec:midpoint-field}); unlike in RVPA, the magnetic field acts in the drifts, which follow the gyration about $\boldsymbol B$ exactly \cite{Patacchini2009}, and the electric field in the kick. The step length varies along the orbit through a time transformation \cite{MikkolaTanikawa1999,PretoTremaine1999,MikkolaAarseth2002} (Section~\ref{subsubsec:adaptive-time-stepping}), and the step is time-symmetric. This structure-preserving perspective is also reflected in geometric particle-in-cell methods for long-time kinetic plasma simulations~\cite{xiaoStructurePreserving2018,xiaoExplicitStructurePreserving2021}. The widely used Boris algorithm provides a familiar example of the long-time advantages associated with volume-preserving particle integration~\cite{qinWhyBorisAlgorithm2013,borisProceedings4thConference1972}.

\begin{figure}[htbp]
    \centering
    \includegraphics[width=0.5\linewidth]{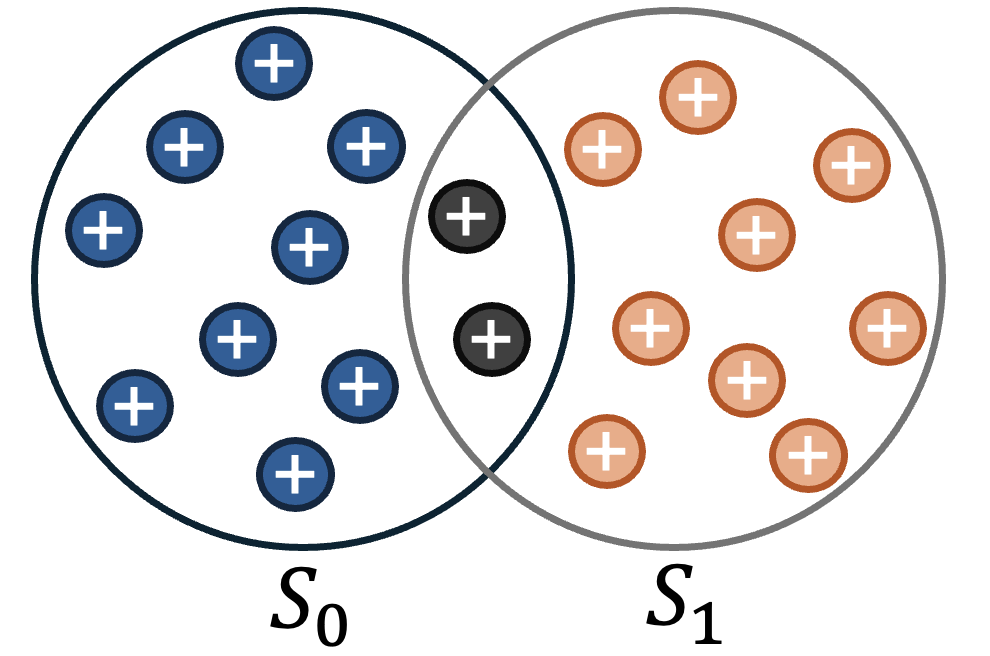}
    \caption{Schematic of interaction-sphere updating during trajectory integration. As the test electron moves, ions in the overlap region are retained, ions left behind leave the sphere, and the ions of the newly entered region are added; they are the same each time the region is entered.}
    \label{fig:debye-incorporation}
\end{figure}

To reduce the cost of direct force summation, the electric field is evaluated only from ions inside a sphere of radius $R_c$ centered on the test electron. As the electron moves, ions in the overlap of the old and new spheres are retained, ions left behind are dropped from the sphere, and the ions of the newly entered region are added, as illustrated in Figure~\ref{fig:debye-incorporation}. The positions of the ions of each region are fixed by a seed tied to that region, so an electron that returns to a region meets the same ions (Algorithm~\ref{ttl:alg}). Although this construction updates the list of ions around the electron at manageable cost, re-centering the sphere introduces a subtle inconsistency between the force-evaluation geometry and the particle update. Section~\ref{subsubsec:midpoint-field} describes the midpoint field evaluation used to suppress the resulting long-time drift.

The disparity between weak long-range interactions and rare close approaches is treated separately through adaptive time stepping. The step size is reduced when the electron approaches an ion and increased in weakly varying regions, allowing close encounters to be resolved without making the remainder of the trajectory unnecessarily expensive. The detailed time-step rule is given in Section~\ref{subsubsec:adaptive-time-stepping}.

The complete method is implemented in the Accurate Particle Tracer (APT) code~\cite{wangAccurateParticleTracer2017}, which provides MPI parallel computing. The electron ensemble is distributed over MPI ranks; each rank advances its assigned particles using the local interaction-sphere force evaluation and adaptive time stepping, and writes their states at common output times. Each electron moves through its own ion background, generated from a global seed and the index of the electron, so the electrons are independent and a trajectory does not depend on the number of ranks. This workflow makes the long-time ensemble calculations practical. Algorithm~\ref{ttl:alg} summarizes the integration of one electron (\ref{ttl:app:proofs}), and a Python implementation of it, which reproduces trajectories of APT to round-off, is provided with the data (see Data availability).

\subsubsection{Midpoint Field Evaluation to Reduce Long-Time Drift}
\label{subsubsec:midpoint-field}

\begin{figure}[htbp]
    \centering
    \includegraphics[width=0.5\linewidth]{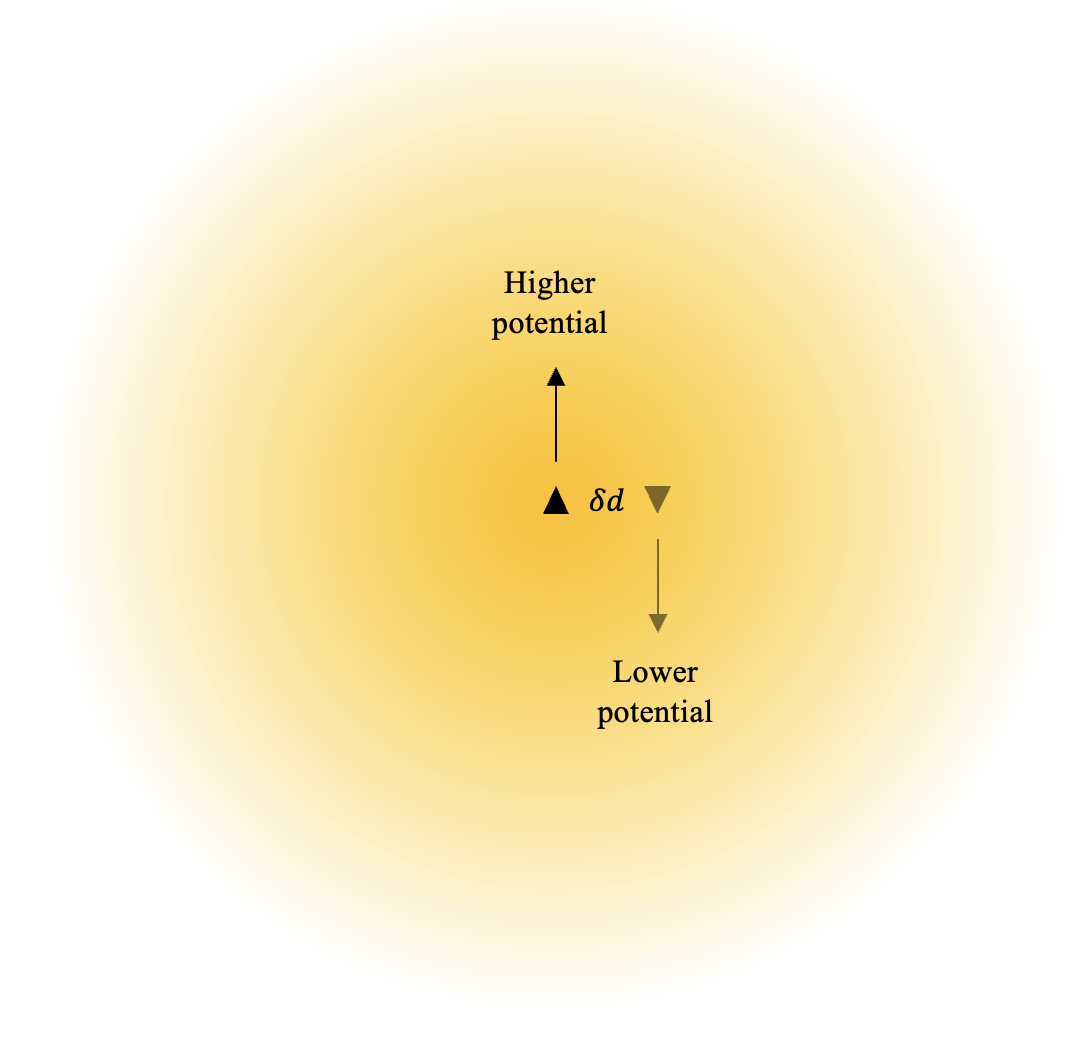}
    \caption{Illustration of the potential mismatch introduced by the moving interaction sphere. Because the electric potential at the sphere center differs from that in the surrounding region, an update with the field at the start of the step can introduce an artificial change in particle energy.}
    \label{fig:potential-change}
\end{figure}

The moving interaction-sphere construction reduces the cost of direct force evaluation, but it also introduces a consistency problem when the field is evaluated only at the beginning of each time step. In that case, the electron is advanced using a field associated with a sphere centered at its previous position, even though the particle moves away from that center during the update. The resulting error is small over a single step but accumulates over long integrations, producing an artificial drift in the particle energy. Because the transport coefficients are inferred from long-time trajectory statistics, this drift can directly contaminate the observables of interest. The geometric origin of this mismatch is illustrated in Figure~\ref{fig:potential-change}.

To restore consistency between the force-evaluation geometry and the particle update, we evaluate the electric field at the temporal midpoint of each step. The procedure is as follows:
\begin{enumerate}
    \item \textbf{Half drift:} starting from the current position $\mathbf{x}^{(n)}$, advance the electron in the magnetic field alone over the first half of the step, to the midpoint $\mathbf{x}^{(n+1/2)}$.

    \item \textbf{Kick (midpoint):} construct the ion set within a sphere of radius $R_c$ centered at $\mathbf{x}^{(n+1/2)}$, compute the midpoint field $\mathbf{E}^{(n+1/2)}$, and change the velocity by the impulse of this field over the step.

    \item \textbf{Update:} complete the advancement from step $n$ to step $n+1$ with a second half drift.
\end{enumerate}

\begin{figure}[htbp]
     \centering
     \begin{subfigure}[b]{\PaperFigurePanelWidth}
         \centering
         \PaperFigureInPanel{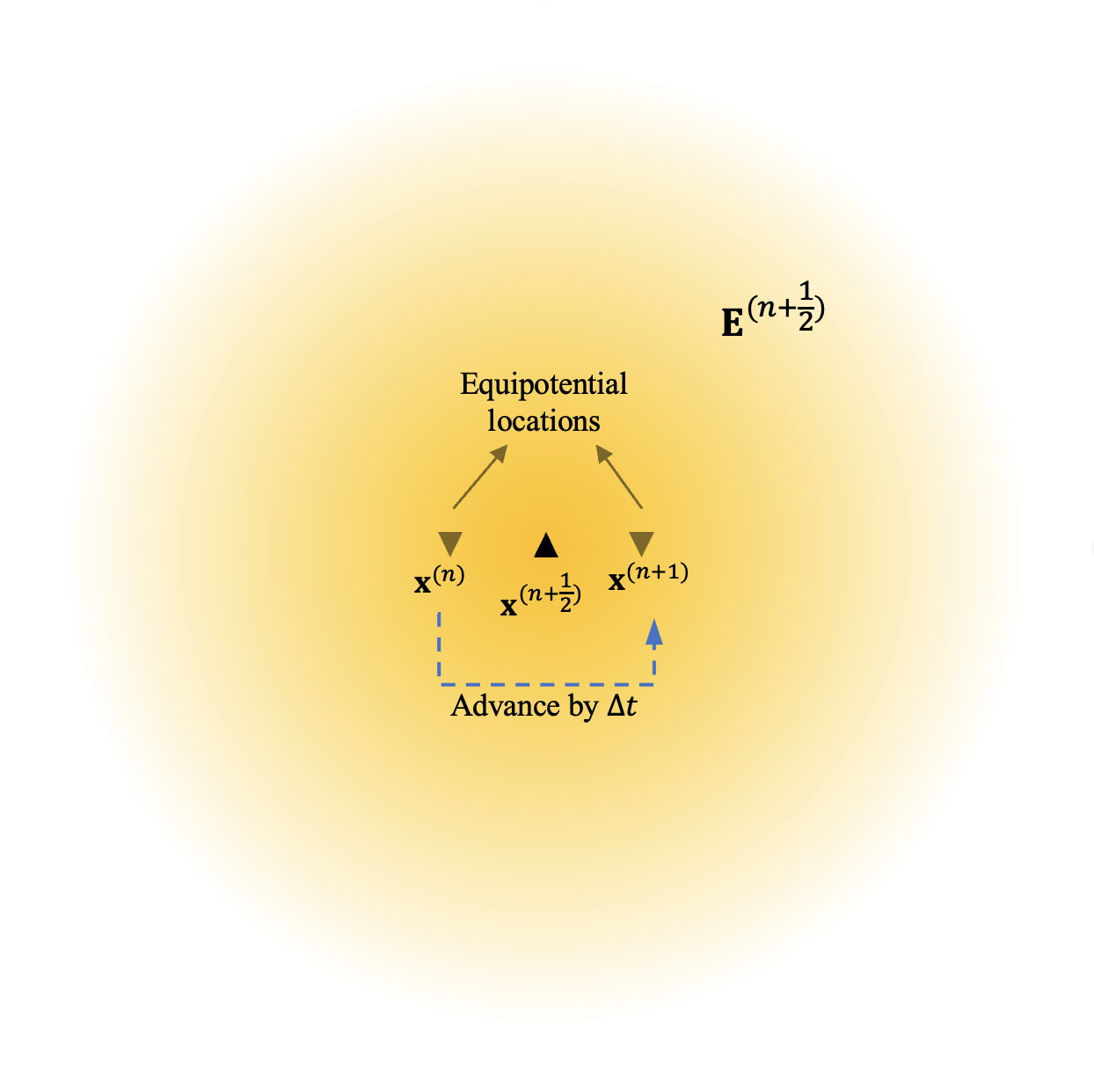}
         \caption{By constructing the midpoint electric field $\mathbf{E}^{(n+1/2)}$, the test electron is advanced between approximately equipotential locations.}
         \label{fig:midpoint-E}
     \end{subfigure}
     \hfill%
     \begin{subfigure}[b]{\PaperFigurePanelWidth}
         \centering
         \PaperFigureInPanel{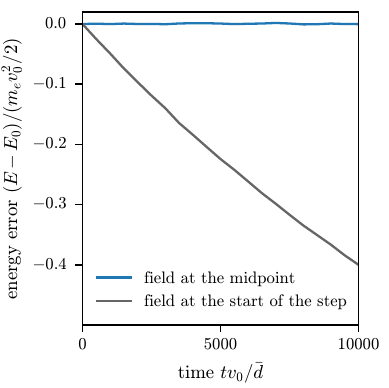}
         \caption{Comparison of long-time energy evolution (mean over 16 electrons, $R_c=0.5\lambda_D$, $\boldsymbol B=0$). The midpoint strategy suppresses the secular energy drift.}
         \label{fig:midpoint-comparison}
     \end{subfigure}
     
     \caption{Midpoint field evaluation strategy and its effect on energy conservation.}
     \label{fig:midpoint-combined}
\end{figure}

This construction evaluates the force at a location representative of the full time step, thereby reducing the mismatch caused by re-centering the sphere. Because the half drift itself brings the electron to the midpoint, the field is evaluated once per step. As illustrated in Figure~\ref{fig:midpoint-E}, the midpoint update advances the particle between approximately comparable potential locations rather than using a field tied entirely to the beginning-of-step geometry. The resulting improvement in long-time behavior is shown in Figure~\ref{fig:midpoint-comparison}: the midpoint scheme largely removes the secular kinetic-energy decrease observed in an update with the field at the start of the step.

\subsubsection{Adaptive Time Stepping for Rare Close Encounters}
\label{subsubsec:adaptive-time-stepping}

\begin{figure}[htbp]
    \centering
    \begin{minipage}[c]{0.31\textwidth}
        {\fontsize{9}{11}\selectfont (a)}\par\smallskip
        \includegraphics[width=\linewidth]{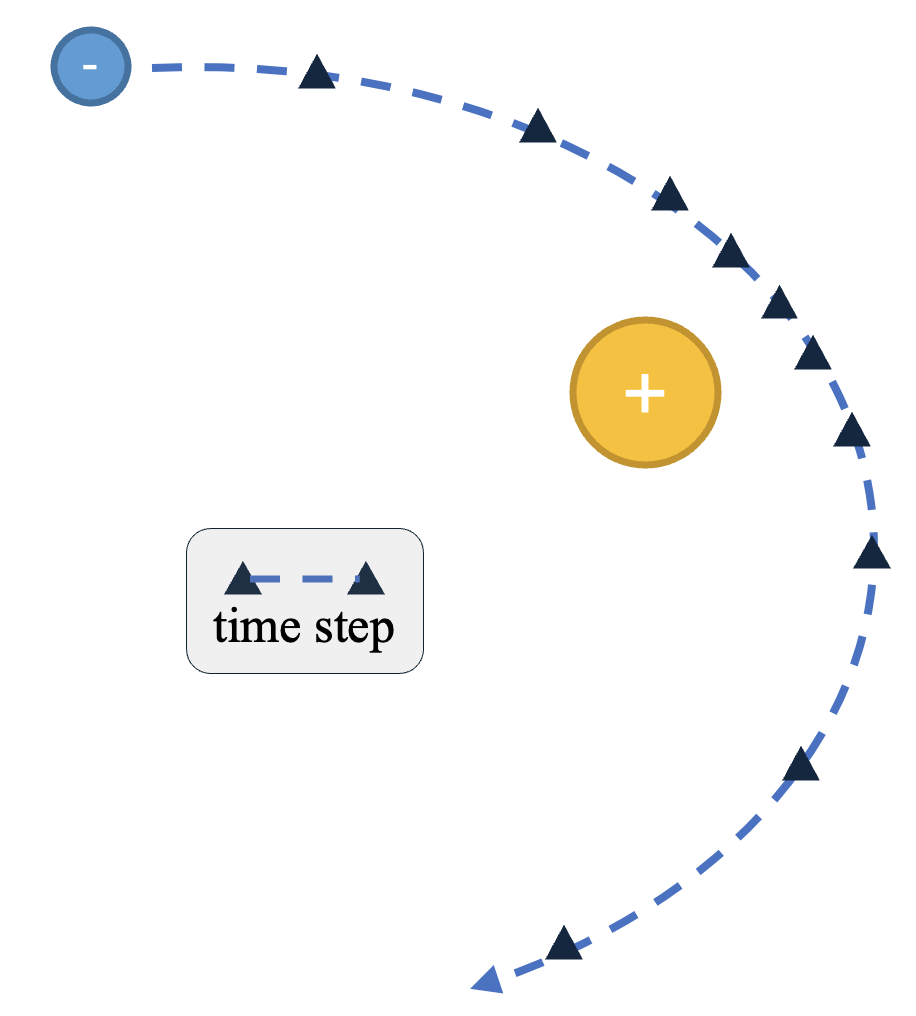}
    \end{minipage}\hfill
    \begin{minipage}[c]{\PaperFigureTwoThirdsWidth}
        \includegraphics[width=\linewidth]{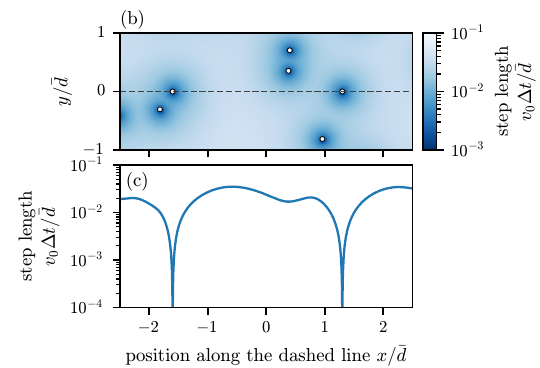}
    \end{minipage}
    \caption{(a) Schematic illustration of adaptive time stepping. Close electron--ion encounters require smaller time steps to resolve the rapidly varying force, whereas weak long-range interactions can be integrated with larger steps. (b) Step length $\Delta t$ of Eq.~\eqref{ttl:eq:g} among ions placed at random in a plane, one per area $\dbar^2$ (dots), for $R_c=3\lambda_D$, $\boldsymbol B=0$ and $\Delta\tilde t=0.05\,\kappa/v_0$; two of the ions lie on the dashed line. (c) Step length along the dashed line of (b). Near each ion the step falls in proportion to the distance to that ion, and it is longest between the ions.}
    \label{fig:adaptive-timestep}
\end{figure}

The wide variation in interaction strength along the trajectory requires a time step that adapts to the local electron--ion separation. Most electron--ion encounters are weak and contribute through the cumulative effect of many small-angle deflections, so they can be integrated accurately with a relatively large time step. Rare close approaches, however, generate rapidly varying forces and require much finer temporal resolution. A fixed time step therefore creates an inefficient compromise: a large step under-resolves close encounters, whereas a uniformly small step makes the weak-field portion of the trajectory unnecessarily expensive (Fig.~\ref{fig:adaptive-timestep}).

We therefore define the time step from the instantaneous distances $r_i$ to the ions within $R_c$, mainly the distance to the nearest one, using these distances as a proxy for the local force-variation scale. The adaptive step is $\Delta t=\Delta\tilde t/g(\mathbf x)$, with a constant $\Delta\tilde t$ and
\begin{equation}
  g(\mathbf x)=\Big(\sum_{r_i<R_c}\psi(r_i)^4\Big)^{1/4}+g_\infty,\qquad
  \psi(r)=\frac{\kappa}{r}\Big(1-\frac{r^2}{R_c^2}\Big)^2,
  \label{ttl:eq:g}
\end{equation}
where $\kappa=\lvert q_iq_e\rvert/(4\pi\epsilon_0)=m_ev_0^2b_0$ and $g_\infty=\Delta\tilde t/\Delta t_{\max}$. Between the ions, where the long-range forcing is weak and varies slowly, the time step is longest; it is bounded by $\Delta t_{\max}=\min(0.05\,\tau_g,\ \dbar/v_0)$, where $\dbar=n^{-1/3}$ is the mean ion spacing. As $r_i$ decreases during a close encounter, $g$ approaches $\kappa/r_i$ and the step, $\Delta t\approx\Delta\tilde t\,r_i/\kappa$, is reduced smoothly in proportion to $r_i$ (Fig.~\ref{fig:adaptive-timestep}b, c), ensuring that the rapidly varying force remains resolved down to head-on passages.

The step length is set through a time transformation \cite{MikkolaTanikawa1999,PretoTremaine1999,MikkolaAarseth2002}: the integration advances a transformed time $\tilde t$, with $\mathrm dt/\mathrm d\tilde t=1/g$, in equal steps $\Delta\tilde t$. An auxiliary variable $W$, updated in the kick, sets the physical duration of the half drifts (\ref{ttl:app:proofs}). Each step of this time-transformed integration is second order and time-symmetric \cite{MikkolaAarseth2002} while its length follows $g$ along the orbit.

Unless otherwise stated, we use $\Delta\tilde t=0.05\,\kappa/v_0$. This parameter controls the fraction of the characteristic length scale traversed during one step and therefore plays a role analogous to a particle-based CFL factor: near an ion, an electron of speed $v_0$ moves $0.05$ of its distance to that ion in one step. The adaptive scheme thus resolves the rare near-field events responsible for large deflections while retaining the efficiency required for long-time ensemble transport calculations.

\subsection{Verification Against Rutherford Scattering}
\label{subsec:rutherford}
Before using the simulation to extract transport coefficients, it is important to verify that the numerical scheme recovers the correct two-body Coulomb-scattering limit. This benchmark is especially relevant here because the long-time transport calculations reported below depend on the accurate treatment of rare close encounters, which are precisely the events most susceptible to numerical error if the integrator or time-step control is inadequate. We therefore isolate a single scattering event and compare the computed deflection angle with the Rutherford prediction. In the Coulomb field of a single ion the deflection angle is known exactly, so this case tests the time-transformed step apart from the ion background.

A stationary ion of charge $+e$ is placed at the origin, its force is the unscreened Coulomb force, and $\boldsymbol B=0$. Without an interaction radius the sum in Eq.~\eqref{ttl:eq:g} reduces to the single term $\kappa/r$, so that $g=\kappa/r+g_\infty$. An electron is launched from $(-\dbar,\,b,\,0)$ with initial speed $v_0$ along the $+x$ direction and impact parameter $b$ between $10^{-8}\,b_0$ and $10^2\,b_0$, and is followed until long after the encounter. In this limit, the theoretical scattering angle is given by the Rutherford formula \cite{rutherfordLXXIXScatteringParticles1911}
\begin{equation}
        \tan \frac{\theta}{2} = \frac{b_0}{b}
        \label{eq:rutherford}
\end{equation}
for an electron that arrives from infinity with speed $v_0$ and impact parameter $b$, where $b_0$ is the impact parameter for $90^\circ$ deflection, Eq.~\eqref{eq:b0}. The computed angle is taken between the asymptotes of the hyperbolic orbits through the initial and the final state, and the Rutherford angle from the orbit through the initial state, so the finite start and end distances do not enter.

\begin{figure}[!htbp]
\centering
\PaperFigureWide{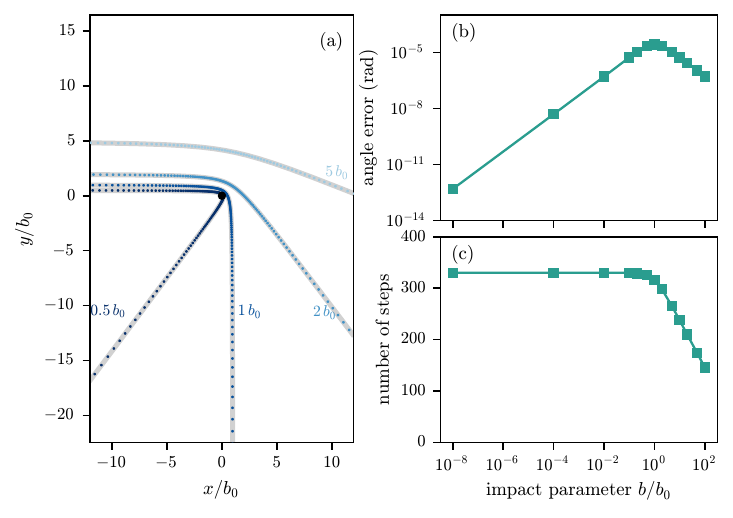}
\caption{Benchmark of the single-collision limit: scattering by a single ion with
the unscreened Coulomb force and $\boldsymbol B=0$. The electron starts at $(-\dbar,\,b,\,0)$ with speed $v_0$ and is
followed for a time $2\dbar/v_0$. (a) Positions after each time-transformed step (dots) for $b=5$, 2,
1 and $0.5\,b_0$, on the analytic Coulomb orbits (gray lines); the ion is at the origin. (b) Error of the deflection
angle and (c) number of steps for the whole path, against the impact parameter $b$, for the time-transformed step with
$\Delta\tilde t=0.05\,\kappa/v_0$. The time-transformed step follows the analytic orbits at every tested impact parameter with
at most 330 steps.}
\label{fig:coulomb-scattering}
\end{figure}

Figure~\ref{fig:coulomb-scattering}(a) shows that the steps lie on the analytic hyperbolic orbits, and the error of the
deflection angle is at most $3\times10^{-5}$\,rad (Fig.~\ref{fig:coulomb-scattering}b). A whole passage takes at most 330 steps at every $b$
(Fig.~\ref{fig:coulomb-scattering}c). This result confirms that the integration scheme reproduces the correct Coulomb-scattering limit, including near-\(90^\circ\) deflections, and therefore provides a reliable foundation for the transport calculations that follow. In particular, it supports the interpretation that any later deviation from reduced collision theory arises from the physics retained in the simulation, rather than from failure to resolve the underlying particle dynamics.

\section{From Trajectories to Transport Benchmarks}
\label{subsec:transport-benchmarks}

The central methodological question is how to compare a trajectory calculation with transport theory, which is usually expressed in terms of reduced coefficients rather than particle paths. In the present model, no collision operator is imposed, so the transport coefficients are extracted from ensemble statistics of the trajectories. To make the dynamics quantitatively comparable with kinetic theory, we introduce two ensemble-level observables extracted from the simulated trajectories: (i) the pitch-angle scattering rate, which is the momentum-relaxation rate in the unmagnetized limit (Section~\ref{subsec:obs-nu}), and (ii) the cross-field diffusion coefficient of the guiding center in the magnetized limit (Section~\ref{subsec:obs-D}). We compare both with an independent-encounter reference computed in the same model; the comparison measures how far the transport runs depart from the assumption of independent encounters (Section~\ref{subsec:ie-reference}). Splitting the guiding-center displacement at reversals of the parallel velocity shows whether an excess of the diffusion coefficient over the reference comes from correlated displacements in successive pieces of the trajectory (Section~\ref{subsec:obs-excursions}). Runs in which the ion configuration is renewed limit how often an electron returns to the same ions, and so measure how much of the excess these returns cause (Section~\ref{subsec:ie-reference}).

Each electron starts at speed $v_0$ in an isotropic random direction in a statistically homogeneous background, so after a short transient the ensemble is stationary. The estimators therefore average over all electrons and over all time origins after the transient, at common output times (Algorithm~\ref{ttl:alg}, Appendix~\ref{app:estimators}). Successive time origins of one electron are correlated, so each uncertainty quoted with $\pm$ is one standard error from a bootstrap over whole electrons \cite{efronBootstrapMethodsAnother1979}, or from a jackknife over batches of passes for the reference (Appendix~\ref{app:ie}).

\subsection{Momentum Relaxation and Pitch-Angle Scattering}\label{subsec:obs-nu}

In the unmagnetized case, the most direct transport signature is the loss of directed momentum (i.e.\ the isotropization) under the cumulative action of screened Coulomb deflections. Because no collision frequency is prescribed in the simulation, the relaxation rate must be inferred from the trajectories themselves. We do so by monitoring the decay of the velocity component along the direction of motion at an earlier time.

Because the electrons start in random directions, each electron's direction of motion $\hat{\mathbf u}=\bv/\lvert\bv\rvert$ is compared with its own direction at an earlier time $s$. The mean over electrons and over all $s$ after the transient defines the autocorrelation function of the direction,
\begin{equation}
\begin{aligned}
C(\tau)&=\Big\langle\underbrace{\hat{\mathbf u}(s+\tau)\cdot\hat{\mathbf u}(s)}_{\makebox[0pt]{\scriptsize cosine of the angle turned in $\tau$}}\Big\rangle &&(\boldsymbol B=0),\\
C(\tau)&=3\,\big\langle\mu(s+\tau)\,\mu(s)\big\rangle &&(\boldsymbol B\neq0).
\end{aligned}
\label{eq:corr}
\end{equation}
In a magnetic field the component of $\hat{\mathbf u}$ perpendicular to $\bb$ rotates at the gyrofrequency without being scattered, so only the pitch-angle cosine $\mu=\hat{\mathbf u}\cdot\bb$, the cosine of the angle between the velocity and $\boldsymbol B$, is used; the factor $3=1/\langle\mu^2\rangle$ of an isotropic ensemble makes $C(0)=1$. Collisions turn the direction in a random walk on the unit sphere, so the mean cosine decays exponentially at a rate called the pitch-angle scattering rate $\nu$ \cite{nanbuTheoryCumulativeSmallangle1997,nanbuMomentumRelaxationCharged1997},
\begin{equation}
C(\tau)\simeq\text{const}\times e^{-\nu\tau},
\label{eq:nu-fit}
\end{equation}
with $\nu$ fitted to $\ln C(\tau)$ by least squares (Appendix~\ref{app:estimators}). Velocity and pitch-angle correlations of this kind have been computed from test-particle trajectories among static scatterers in a magnetic field \cite{kuzmanyMagnetotransportTwodimensionalLorentz1998} and in magnetic turbulence \cite{fraschettiEarlytimeVelocityAutocorrelation2012}.

This observable is chosen because it provides a direct bridge between microscopic dynamics and the classical Lorentz-picture description of momentum relaxation. For scattering by static ions it decays at the standard electron--ion collision frequency of binary-collision theory \cite{chandrasekhar1943stochastic}, \cite[Sec.~3.2]{helanderCollisionalTransportMagnetized2002},
\begin{equation}
\nu_{ei}(v)=\frac{n_{i} q_{i}^{2} q_e^{2}}{4 \pi m_{e}^{2} \epsilon_{0}^{2} v^{3}} \ln \Lambda,
\label{eq:nu_classical_results}
\end{equation}
where $\ln\Lambda$ is the Coulomb logarithm. Following Liboff~\cite{liboff1959} and Escande et al.~\cite{escandeUniformDerivationCoulomb2015}, who summed independent binary encounters in the Debye-screened field, we sum them for the screened field of Eq.~\eqref{eq:debye-electric} truncated at $R_c=c\lambda_D$ and obtain
\begin{equation}
\ln\Lambda_{\rm eff}=\ln\frac{\lambda_D}{b_0}+K(c),
\label{eq:lnL-eff}
\end{equation}
where the constant $K(c)$ accounts for the screening and the truncation at $R_c$ (Appendix~\ref{app:coulomb-log}). We write $\nu_0$ for Eq.~\eqref{eq:nu_classical_results} at $v=v_0$ with $\ln\Lambda=\ln\Lambda_{\rm eff}$; it is the unmagnetized reference of Section~\ref{subsec:unmagnetized_nu} and, with $c=3$, the rate unit of Section~\ref{subsec:magnetized_diffusion}.

\subsection{Cross-Field Diffusion in the Magnetized Limit}\label{subsec:obs-D}

In the magnetized case, transport is not adequately characterized by momentum relaxation alone. The more relevant quantity is the slow irreversible spreading of particle trajectories across the magnetic field, after the fast gyromotion has been averaged out. The corresponding observable is the long-time diffusion coefficient of the guiding center, extracted from its mean-squared displacement (MSD).

The electron gyrates about a guiding center, the center of the gyration,
\begin{equation}
\mathbf X=\bx-\frac{\bv\times\bb}{\Omega} ,
\label{eq:gc}
\end{equation}
where $\Omega=eB/m_e$ is the cyclotron frequency (Section~\ref{subsec:model}). The guiding center moves along $\boldsymbol B$. The magnetic field only rotates the velocity, which does not move $\mathbf X$ across the field, so $\mathbf X$ moves across the field only when an ion changes the velocity. A velocity change $\Delta\bv$ moves it by $-\Delta\bv\times\bb/\Omega$, a distance $\lvert\Delta\bv_\perp\rvert/\Omega$, where $\Delta\bv_\perp$ is the component of $\Delta\bv$ perpendicular to $\boldsymbol B$ \cite{longmireDiffusionChargedParticles1956,helanderCollisionalTransportMagnetized2002}.

With $\mathbf X_\perp$ the component of $\mathbf X$ perpendicular to $\boldsymbol B$, the cross-field diffusion coefficient is the growth rate of the mean-squared displacement \cite{taylorDiffusionContinuousMovements1922},
\begin{equation}
D_\perp=\frac14\,\frac{d}{d\tau}\Big\langle\big\lvert\mathbf X_\perp(s+\tau)-\mathbf X_\perp(s)\big\rvert^2\Big\rangle,
\label{eq:Dperp}
\end{equation}
fitted as the slope from one scattering time $1/\nu_1$ to the end of the run, where $\nu_1$ is $\nu_0$ with the encounters at impact parameters above $\rho_v$ removed from the Coulomb logarithm \cite{montgomeryMagneticFieldDependence1974} (Table~\ref{tab:runs}, Appendix~\ref{app:estimators}). Running diffusion coefficients of this type are used, for example, in test-particle studies of turbulent transport \cite{naulinDispersionIdealParticles1999}.

\begin{figure}[!htbp]
  \centering
  \PaperFigureWide{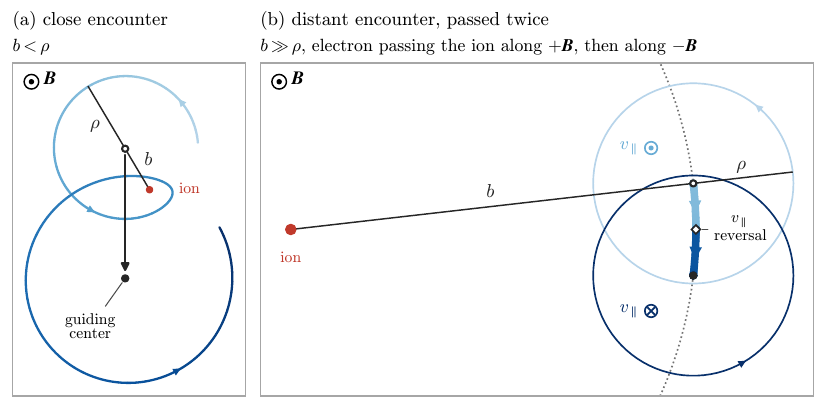}
  \caption{Close and distant encounters of an electron with one ion in a magnetic field. A close encounter makes the guiding center jump; a distant one makes it drift around the ion the same way on either pass, so the displacements of two passes add. The encounters are viewed along $\boldsymbol B$ (out of the page) and computed for the screened field of Eq.~\eqref{eq:debye-electric}; the $90^\circ$ impact parameter $b_0$ of Eq.~\eqref{eq:b0} is enlarged so that the drift is visible. Colors run from light to dark with time; open circle: initial guiding center, filled circle: final guiding center; $b$: impact parameter, measured from the guiding center; $\rho$: gyroradius. (a) Close encounter, $b<\rho$: the ion turns the velocity, $\mu$ and with it the gyroradius change; arrow: jump of the guiding center. (b) Distant encounter, $b\gg\rho$ ($b$ not to scale), passed twice: the electron passes the ion along $+\boldsymbol B$ ($\odot\,v_\parallel$), reverses (diamond), and passes it again along $-\boldsymbol B$ ($\otimes\,v_\parallel$). Thin circles: gyro-orbit at the start and at the end; thick line: guiding center averaged over a gyroperiod. Dotted: circle about the ion.}
  \label{fig:close-distant}
\end{figure}

An ion moves the guiding center in one of two ways, depending on the impact parameter $b$ of the encounter, measured from the guiding center, compared with the gyroradius $\rho\le\rho_v=v_0/\Omega$ of the electron (Fig.~\ref{fig:close-distant}).

Close encounters, with $b\lesssim\rho$, make the guiding center jump (Fig.~\ref{fig:close-distant}a). The ion turns the velocity, so $\mu$, the gyroradius and the gyrophase change, and the guiding center jumps by $\lvert\Delta\bv_\perp\rvert/\Omega$. A return comes at an unrelated gyrophase and rarely comes as close again, so it rarely repeats the jump (Fig.~\ref{fig:weak-strong}a). These jumps give the classical diffusion, in which the guiding center moves by about a gyroradius per scattering time; under isotropic pitch-angle scattering at the rate $\nu$, averaged over an isotropic distribution of $\mu$, they give $D_\perp=\nu v_0^2/(3\Omega^2)$ \cite{abelLinearizedModelFokkerPlanck2008}.

Distant encounters, with $b\gg\rho$, make the guiding center drift (Fig.~\ref{fig:close-distant}b). The guiding center moves with the $\mathbf E\times\mathbf B$ drift around the ion, and the pitch-angle cosine $\mu$ returns to its initial value. We call the passage of the electron through the sphere of radius $R_c$ about the ion a pass. The displacement of a pass grows as $\lvert\mu\rvert$ decreases and does not depend on the sign of $v_\parallel$ \cite{dubinTestParticleDiffusion1997}. An electron with small $\lvert\mu\rvert$, which moves nearly perpendicular to $\boldsymbol B$, passes the same ion again when an encounter or the field of the other ions reverses its parallel velocity $v_\parallel=\mu v$. The return pass then displaces the guiding center the same way as the first, so the two displacements are not independent.

Both kinds of encounter are known from earlier work. Close encounters are the random walk of classical transport theory \cite{longmireDiffusionChargedParticles1956}, and the magnetic field corrects their deflection angle \cite{jiangTransverseRutherfordScattering2022}. Distant encounters give the long-range diffusion, computed by integrating the drift along unperturbed orbits and counting each encounter once \cite{oneilNewTheoryTransport1985}, and measured in pure ion plasmas \cite{andereggTestParticleTransport1997}. Dubin included the repeated collisions of the same pair of guiding centers in a single-species plasma, where both particles move along the field \cite{dubinTestParticleDiffusion1997}.

In this paper we measure the effect of repeated passes by comparing $D_\perp$ with its value $D_\perp^{\rm IE}$ for independent encounters in the same model (Section~\ref{subsec:ie-reference}). The ratio $D_\perp/D_\perp^{\rm IE}$ is 1 for independent encounters. Section~\ref{subsec:obs-excursions} describes how we locate the repeated passes. We quote diffusion coefficients in units of $D_0=\nu_0v_0^2/(3\Omega^2)$, the classical diffusion at the unmagnetized rate $\nu_0$; in these units a coefficient proportional to $B^{-2}$ is constant.

\subsection{Three Calculations That Isolate the Repeated Passes}\label{subsec:ie-reference}

\begin{figure}[!htbp]
  \centering
  \PaperSchematic{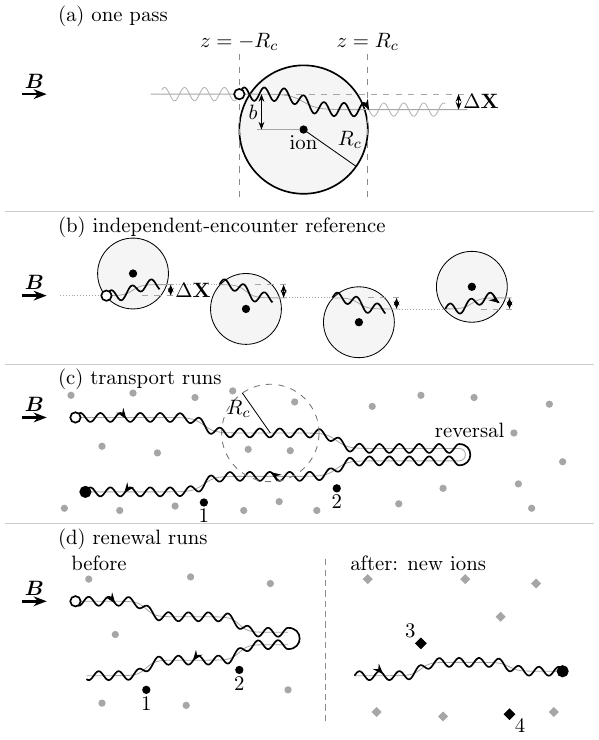}
  \caption{The three calculations. The electron meets a new ion at every pass in the reference, passes the same ions at every return in the transport runs, and does so in the renewal runs only until the ions are renewed. Side views in the plane that contains $\boldsymbol B$; not to scale. Thick line: helix of the electron; thin gray line: path of its guiding center; open circle: start, filled circle: end; dashed gray line: path of the guiding center before a shift. (a) One pass of the reference: the guiding center passes the ion at impact parameter $b$, from $z=-R_c$ until $\lvert z\rvert$ first exceeds $R_c$, and the pass shifts it by $\Delta\mathbf X$ (drawn in the page for visibility; a distant pass shifts it around the ion, Fig.~\ref{fig:close-distant}). (b) Independent-encounter reference: the shifts of successive passes have random directions and sizes and are summed as independent events, Eq.~\eqref{eq:nuIE}. (c) Transport runs: all ions within $R_c$ act on the electron at once (simultaneous interactions, dashed circle); after a reversal the electron passes ions 1 and 2 again, and the guiding center is shifted the same way. (d) Renewal runs, before and after a renewal: ions 1 and 2 shift the guiding center the same way on both passes, as in (c); the new ions 3 and 4 (diamonds) shift it in unrelated directions. In (c) and (d) the two passes are drawn slightly apart; the gap is not a shift.}
  \label{fig:ie-schematic}
\end{figure}

We compare three calculations (Table~\ref{tab:model-comparison}, Fig.~\ref{fig:ie-schematic}). First, in the transport runs all ions within $R_c$ act on the electron at once (simultaneous interactions), and the ions stay in place, so an electron that reverses its parallel velocity passes the same ions again (repeated passes). Second, the independent-encounter reference has neither: one ion acts at a time, and every pass meets a new ion, as if the ions were redistributed at random after each pass. This is the Boltzmann description of scattering by static centers \cite{dmitrievAnomalousLowFieldClassical2002,dmitrievNonBoltzmannClassicalCorrection2005}. Third, the renewal runs keep the simultaneous interactions and limit the repeated passes to about two.
\begin{table}[!htbp]
    \centering
    \caption{The three calculations compared in Section~\ref{sec:results}. They differ in the simultaneous interactions and in the repeated passes.}
    \label{tab:model-comparison}
    \small
    \begin{tabular}{@{}>{\raggedright\arraybackslash}p{0.17\textwidth}>{\raggedright\arraybackslash}p{0.25\textwidth}>{\raggedright\arraybackslash}p{0.24\textwidth}>{\raggedright\arraybackslash}p{0.24\textwidth}@{}}
      \toprule
       & Independent-encounter reference & Renewal runs & Transport runs \\
      \midrule
      Potential & \multicolumn{3}{>{\raggedright\arraybackslash}p{0.75\textwidth}@{}}{Debye-screened potential, Eq.~\eqref{eq:debye-electric}, truncated at $R_c$ (all three)} \\
      \addlinespace
      Simultaneous interactions & None: one ion at a time & All ions with $r<R_c$ & All ions with $r<R_c$ \\
      \addlinespace
      Repeated passes & None: each pass meets a new ion, and the rates of all passes are summed, Eq.~\eqref{eq:nuIE} & About two: the ions are renewed every $\tau_c=0.03/\nu_1$ & As often as the orbit returns: the ions stay in place \\
      \addlinespace
      Orbit & $\boldsymbol B=0$: binary collision summarized by $\ln\Lambda_{\rm eff}$, Eq.~\eqref{eq:lnL-eff}; $\boldsymbol B\neq0$: integrated with the step of Section~\ref{subsec:numerical-method} & \multicolumn{2}{>{\raggedright\arraybackslash}p{0.49\textwidth}@{}}{Integrated with the step of Section~\ref{subsec:numerical-method} in the superposed field, Eq.~\eqref{eq:superposed-field}} \\
      \addlinespace
      Fields & $B=0$; $10^{-4}$ to $5\times10^{-2}\,\mathrm T$ & $10^{-2}\,\mathrm T$ & $B=0$; $10^{-4}$ to $5\times10^{-2}\,\mathrm T$ \\
      \bottomrule
    \end{tabular}
\end{table}

Comparing the three calculations separates the two possible causes of the excess. First, the transport runs against the reference give the whole excess. Second, the transport runs against the renewal runs give the part that comes from passing the same ions again; the renewal runs and the reference still differ in both respects, so comparing these two separates neither. Third, the simultaneous interactions alone are tested without a magnetic field, where they add up to independent binary encounters \cite{escandeUniformDerivationCoulomb2015}: the transport runs reproduce the binary-collision rate $\nu_0$ (Section~\ref{subsec:unmagnetized_nu}).

The renewal runs limit the repeated passes by replacing the ion configuration with a new random one every $\tau_c=0.03/\nu_1$, a few hundredths of a scattering time, in runs that repeat the transport runs at $10^{-2}\,\mathrm T$ for the same electrons (Fig.~\ref{fig:ie-schematic}d). An electron reverses about once in $\tau_c$ ($\mu$ changes sign 1.2 times per $\tau_c$ on average), so it passes the same ions about twice before they are renewed, against once in the reference and as often as the orbit returns in the transport runs.

We compute the independent-encounter reference from simulated passes of an electron by a single ion, because analytic results for a pass exist only in limits, for an electron moving across the field \cite{jiangTransverseRutherfordScattering2022} and for distant passes on straight paths \cite{dubinTestParticleDiffusion1997}. The reference gives the values of $\nu$ and $D_\perp$ that independent encounters would give for the same ion field, interaction radius and magnetic field. The passes are integrated with the step of Section~\ref{subsec:numerical-method} and the parameters of the transport runs, so the reference includes both close and distant encounters. Each pass follows the orbit in the field of the ion, as in classical-trajectory Monte Carlo \cite{nersisyanBinaryCollisionsCharged2009} and the generalized Boltzmann theory \cite{joseGeneralizedBoltzmannKinetic2020}, and unlike integration along unperturbed orbits \cite{dubinTestParticleDiffusion1997}.

A pass (Fig.~\ref{fig:ie-schematic}a) starts at $z=-R_c$, with the ion at the origin and $\boldsymbol B$ along $z$, and the electron gyrating about a field line at impact parameter $b$ from the ion. It ends when $\lvert z\rvert$ first exceeds $R_c$ again, having changed the pitch-angle cosine by $\Delta\mu$ and displaced the guiding center by $\Delta\mathbf X$. Per unit time, an electron with pitch-angle cosine $\mu$ passes $n\mu v_0\,d^2b$ ions whose distance from its field line lies in $d^2b$. The reference treats every pass as an independent event, so $\nu$ and $D_\perp$ follow as for a random walk: the mean squared change per pass times the rate of such passes. We integrate $M$ passes. The $j$th pass starts with $\mu_j$, impact parameter $\mathbf b_j$ (a vector across $\boldsymbol B$) and gyrophase $\zeta_j$, drawn at random with a probability density $p_j$ that favors passes near the ion, and we sum
\begin{equation}
\nu^{\rm IE}=\frac32\,nv_0\,\frac1M\sum_{j=1}^{M}\frac{\mu_j\,\Delta\mu_j^2}{p_j},\qquad
D_\perp^{\rm IE}=\frac{nv_0}{4}\,\frac1M\sum_{j=1}^{M}\frac{\mu_j\,\lvert\Delta\mathbf X_j\rvert^2}{p_j}.
\label{eq:nuIE}
\end{equation}
Each term is the squared change in one pass, weighted by how often an electron makes such a pass. The density $p_j$ refers to the measure $d\mu\,d^2b\,d\zeta/(2\pi)$, so the sums estimate the integrals of the squared changes over the flux of passes and over an isotropic ensemble.

Following Dubin \cite{dubinTestParticleDiffusion1997}, we drop the drift of slow passes, those with small $\lvert\mu\rvert$, which scattering by the other ions ends early; this cutoff does not change our conclusions (Appendix~\ref{app:ie}).

\subsection{Excursion Decomposition}\label{subsec:obs-excursions}

To test whether returns to the same ions push the guiding center the same way, we cut each trajectory where the parallel velocity reverses, that is, where $\mu$ changes sign. The pieces are excursions, numbered $k=1,2,\dots$ in time; during one excursion the electron moves one way along $\boldsymbol B$. Excursion $k$ of electron $i$ moves the guiding center across the field by $\Delta\mathbf G_{i,k}$. For each electron the squared displacement over the run splits into single-excursion terms and cross terms between pairs of excursions \cite{taylorDiffusionContinuousMovements1922}, so that
\begin{equation}
\begin{aligned}
D_\perp&\approx\frac{1}{4Nt_{\rm end}}\sum_{i=1}^{N}\big\lvert\mathbf X_{\perp,i}(t_{\rm end})-\mathbf X_{\perp,i}(0)\big\rvert^2\\
&=\underbrace{\frac{1}{4Nt_{\rm end}}\sum_{i=1}^{N}\sum_{k}\lvert\Delta\mathbf G_{i,k}\rvert^2}_{\text{single-excursion terms}}
+\underbrace{\frac{1}{4Nt_{\rm end}}\sum_{i=1}^{N}2\sum_{k<l}\Delta\mathbf G_{i,k}\cdot\Delta\mathbf G_{i,l}}_{\text{cross terms}},\\
\Delta\mathbf G_{i,k}&=\mathbf X_{\perp,i}(t_{i,k})-\mathbf X_{\perp,i}(t_{i,k-1}),
\end{aligned}
\label{eq:excursions}
\end{equation}
where $N$ is the number of electrons, all followed for the same time $t_{\rm end}$, $t_{i,1},\dots,t_{i,K_i}$ are the times of the reversals of $v_\parallel$ that end an excursion of electron $i$, $t_{i,0}=0$ and $t_{i,K_i+1}=t_{\rm end}$.

The first sum is what independent excursions give: if successive excursions moved the guiding center in unrelated directions, the cross terms would average to zero and only this sum would remain. The second sum, the cross terms, measures whether different excursions push the guiding center the same way: a return that passes the same ions does so and makes the cross terms positive. Figure~\ref{fig:excursions}a gives the cross terms and the excess $D_\perp/D_\perp^{\rm IE}-1$ with $D_\perp$ from the first line of Eq.~\eqref{eq:excursions}, all divided by $D_\perp^{\rm IE}$.

A reversal is counted when $\mu$ has passed $10^{-3}$ beyond zero, so that small oscillations of $\mu$ about zero are not counted. A reversal inside the potential well of an ion, where $\lvert\bv\rvert$ differs from $v_0$ by $1\%$ or more, belongs to a pass and does not end an excursion. The uncertainties come from a bootstrap over electrons with 2000 resamples.

\begin{table}[!htbp]
    \centering
    \caption{Parameters of the transport runs and of the independent-encounter reference. The rate $\nu_1$ is $\nu_0$ with the encounters at impact parameters above $\rho_v$ removed from the Coulomb logarithm \cite{montgomeryMagneticFieldDependence1974}; it sets the run lengths and the renewal interval $\tau_c$. We call $1/\nu_0$ the collision time and $1/\nu_1$ the scattering time.}
    \label{tab:runs}
    \footnotesize
    \setlength{\tabcolsep}{1pt}
    {\footnotesize\raggedright
    Common to the runs: $n=10^{12}\,\mathrm{m^{-3}}$, $T_e=100\,\mathrm K$; $\dbar=1.0\times10^{-4}\,\mathrm m=1.8\times10^3\,b_0$, $\lambda_D=6.90\,\dbar$, $b_0=5.6\times10^{-8}\,\mathrm m$; electron speed $v_0=6.74\times10^4\,\mathrm{m\,s^{-1}}$, direction isotropic, start at a random point, each electron in its own static ion background (Algorithm~\ref{ttl:alg}); $\Delta\tilde t=0.05\,\kappa/v_0$, $\Delta t_{\max}=\min(0.05\,\tau_g,\ \dbar/v_0)$, $\ell=0.5\,\dbar$; output every $0.01/\nu_0$.\par}
    \smallskip
    \begin{tabular*}{\textwidth}{@{\extracolsep{\fill}}lcclcl@{}}
      \toprule
      Calculation & $B$ (T) & $c$ & Electrons & Length & Used in \\
      \midrule
      Unmagnetized & 0 & 0.5, 1, 3 & 2048 each & $5/\nu_0$ & Section~\ref{subsec:unmagnetized_nu} \\
      Magnetized & $10^{-4}$--$5\times10^{-2}$ & 3 & 256 (512 at $10^{-3}$) & $6/\nu_1$ & Section~\ref{subsec:magnetized_diffusion} \\
      Long run & $10^{-2}$ & 3 & 247 & $40/\nu_1$ & Fig.~\ref{fig:vz-vperp-evolution} \\
      Renewal & $10^{-2}$ & 3 & 128 (127 used) & $6/\nu_1$ & Section~\ref{subsec:excess-origin} \\
      Reference & $10^{-4}$--$5\times10^{-2}$ & 3 & $1.2\times10^6$--$2\times10^7$ passes & -- & Section~\ref{subsec:ie-reference} \\
      \bottomrule
    \end{tabular*}

    \smallskip
    {\footnotesize\raggedright
    Magnetized runs: $\boldsymbol B=B\hat{\boldsymbol z}$ with $B=10^{-4}$, $10^{-3}$, $3\times10^{-3}$, $10^{-2}$ and $5\times10^{-2}\,\mathrm T$; $\nu_1/\nu_0=1.000$, 0.951, 0.852, 0.723 and 0.546. Renewal runs: the first 128 electrons of the magnetized runs at $10^{-2}\,\mathrm T$, of which 127 are used, the stopped electron excluded (Table~\ref{tab:estimators}); ions renewed every $\tau_c=0.03/\nu_1$ (Section~\ref{subsec:ie-reference}), with the speed of the electron reset so that the total energy is unchanged. Reference: same plasma and step; $\mu\ge10^{-3}$; $2\times10^6$, $2\times10^7$, $8\times10^6$, $10^7$ and $1.2\times10^6$ passes at the five fields; cutoff of slow passes in Appendix~\ref{app:ie}.\par}
\end{table}

\section{Collisional Transport Benchmarks}
\label{sec:results}

This section uses the trajectory-based observables introduced in Section~\ref{subsec:transport-benchmarks} to benchmark the independent-encounter description against direct Newton--Lorentz dynamics in the same screened ion background. We first consider momentum relaxation in the unmagnetized limit (Section~\ref{subsec:unmagnetized_nu}) and then compare pitch-angle scattering and cross-field diffusion in the magnetized limit with the independent-encounter reference (Section~\ref{subsec:magnetized_diffusion}). Section~\ref{subsec:excess-origin} locates the origin of the excess of the cross-field diffusion over this reference with the excursion decomposition and the renewal of the ion configuration. All runs use the weakly coupled plasma of Table~\ref{tab:runs}, $n=10^{12}\,\mathrm{m^{-3}}$ and $T_e=100\,\mathrm K$, with $N_D=1.4\times10^3$ ions in a Debye sphere, and the step parameters of Section~\ref{subsec:numerical-method}.

\subsection{Unmagnetized momentum relaxation: comparison with classical $\nu_{ei}$}
\label{subsec:unmagnetized_nu}

In the unmagnetized limit, the first benchmark is whether momentum relaxation emerging from direct Coulomb forcing reproduces the rate predicted by reduced kinetic theory. Without a magnetic field, successive encounters of an electron are with different ions, and the scattering rate should equal the binary-collision rate $\nu_0$ of Eq.~\eqref{eq:nu_classical_results} for the same truncated potential. To isolate this comparison, we fix the density and temperature, set $B=0$, and vary the interaction radius over $c=0.5$, 1 and 3 (Tables~\ref{tab:runs} and~\ref{tab:unmag}); the rate is fitted as in Eq.~\eqref{eq:nu-fit}.

\begin{figure}[!htbp]
  \centering
  \PaperFigureFull{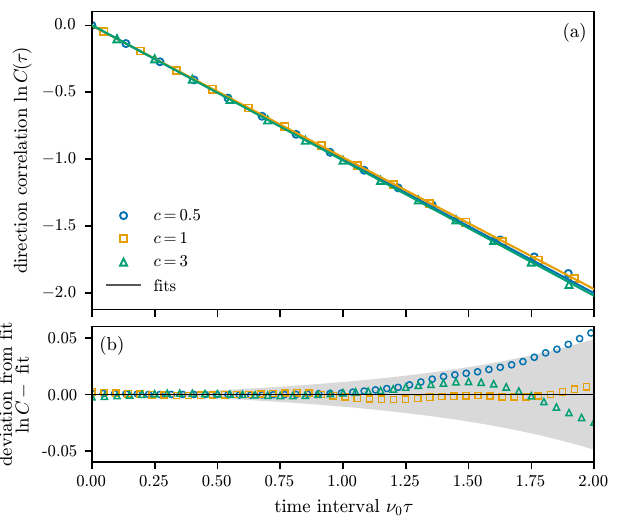}
  \caption{Extraction of the momentum-relaxation rate in the unmagnetized case, for $c=0.5$, 1 and 3. (a) $\ln C(\tau)$ of Eq.~\eqref{eq:corr} against $\nu_0\tau$ (markers), with the exponential fits of Eq.~\eqref{eq:nu-fit} (lines, in the color of each radius). (b) Residual of $\ln C$ from the fit; gray band: bootstrap standard error (the largest of the three radii). $\ln C$ decays linearly at all three radii, so a single rate describes the relaxation, and the fitted rates agree with $\nu_0$ (Table~\ref{tab:unmag}).}
  \label{fig:unmag-corr}
\end{figure}

Figure~\ref{fig:unmag-corr} shows that $\ln C(\tau)$ is linear at all three radii. Table~\ref{tab:unmag} lists the fitted rates. They agree with $\nu_0$ to within $1.6\%$ at each radius, and the inverse-variance weighted mean is $\nu/\nu_0=0.998\pm0.007$.

\begin{table}[!htbp]
    \centering
    \caption{Pitch-angle scattering rate without a magnetic field, for three interaction radii $R_c=c\lambda_D$. $N_c$ is the mean number of ions in the sphere, $\ln\Lambda_{\rm eff}$ is Eq.~\eqref{eq:lnL-eff}, and the uncertainties are bootstrap standard errors over electrons. The weighted mean of the three rates, $\nu/\nu_0=0.998\pm0.007$, has $\chi^2=2.9$ for 2 degrees of freedom.}
    \label{tab:unmag}
    \small
    \begin{tabular}{cccc}
      \toprule
      $c$ & $N_c$ & $\ln\Lambda_{\rm eff}$ & $\nu/\nu_0$ \\
      \midrule
      0.5 & 172 & 8.152 & $1.000\pm0.012$ \\
      1 & $1.4\times10^3$ & 8.671 & $0.984\pm0.011$ \\
      3 & $3.7\times10^4$ & 9.023 & $1.011\pm0.012$ \\
      \bottomrule
    \end{tabular}
\end{table}

\subsection{Magnetized transport: scattering rate and cross-field diffusion}
\label{subsec:magnetized_diffusion}

In the magnetized limit, the relevant transport benchmark is the irreversible spreading of trajectories across the magnetic field. The question addressed here is whether independent encounters still give the pitch-angle scattering rate and the diffusion coefficient inferred directly from resolved trajectories. We compare both observables with the independent-encounter reference of Section~\ref{subsec:ie-reference} at five fields, from $\rho_v=5.6\lambda_D$ ($10^{-4}\,\mathrm T$) to $0.011\lambda_D$ ($5\times10^{-2}\,\mathrm T$), with $c=3$ (Table~\ref{tab:magnetized}).

\begin{table}[!htbp]
    \centering
    \caption{Magnetized runs at $c=3$. Rates are in units of $\nu_0$ and diffusion coefficients in units of $D_0=\nu_0v_0^2/(3\Omega^2)$. Transport runs: $\nu$ and $D_\perp$, Eqs.~\eqref{eq:nu-fit} and~\eqref{eq:Dperp}; independent-encounter reference: $\nu^{\rm IE}$ and $D_\perp^{\rm IE}$, Eq.~\eqref{eq:nuIE}. Uncertainties are bootstrap standard errors over electrons, combined with the sampling error of the reference. The numbers of electrons are given in Table~\ref{tab:runs}.}
    \label{tab:magnetized}
    \footnotesize
    \setlength{\tabcolsep}{4pt}
    \begin{tabular}{@{}lccccc@{}}
      \toprule
      $B$ (T) & $10^{-4}$ & $10^{-3}$ & $3\times10^{-3}$ & $10^{-2}$ & $5\times10^{-2}$ \\
      $\rho_v/\lambda_D$ & 5.56 & 0.556 & 0.185 & 0.0556 & 0.0111 \\
      \midrule
      \multicolumn{6}{@{}l}{Scattering rate} \\
      \quad reference, $\nu^{\rm IE}/\nu_0$ & 0.993 & 0.876 & 0.772 & 0.650 & 0.503 \\
      \quad ratio, $\nu/\nu^{\rm IE}$ & $0.91\pm0.05$ & $1.06\pm0.04$ & $1.05\pm0.07$ & $1.08\pm0.06$ & $1.10\pm0.07$ \\
      \midrule
      \multicolumn{6}{@{}l}{Cross-field diffusion} \\
      \quad transport runs, $D_\perp$ & $1.16\pm0.09$ & $1.83\pm0.10$ & $4.00\pm0.33$ & $10.2\pm0.9$ & $46.6\pm5.8$ \\
      \quad reference, $D_\perp^{\rm IE}$ & 1.06 & 1.44 & 1.91 & 2.33 & 2.78 \\
      \quad ratio, $D_\perp/D_\perp^{\rm IE}$ & $1.09\pm0.08$ & $1.28\pm0.08$ & $2.10\pm0.20$ & $4.4\pm0.4$ & $16.7\pm2.1$ \\
      \bottomrule
    \end{tabular}
\end{table}

\subsubsection{Pitch-angle scattering rate}

\begin{figure}[!htbp]
  \centering
  \PaperFigureWide{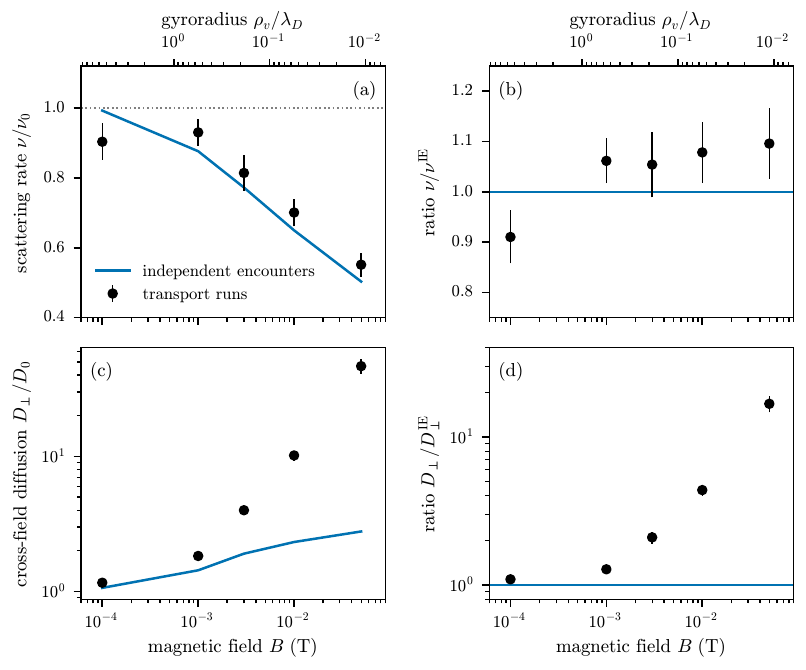}
  \caption{Pitch-angle scattering rate and cross-field diffusion against the magnetic field, at $c=3$; the top axis gives the gyroradius $\rho_v=v_0/\Omega$ in units of the Debye length. Independent encounters give the scattering rate to within $10\%$ at every field, but they give the cross-field diffusion only at $10^{-4}\,\mathrm T$ and underestimate it by a factor that grows with the field, to 17 at $5\times10^{-2}\,\mathrm T$ (Table~\ref{tab:magnetized}). (a) Rate in units of the unmagnetized rate $\nu_0$: transport runs (markers) and independent-encounter reference (line); dotted: $\nu=\nu_0$. (b) Ratio $\nu/\nu^{\rm IE}$; line: 1. (c) Cross-field diffusion coefficient of the transport runs (markers) and of the reference (line), in units of $D_0$ of each field, in which a coefficient proportional to $B^{-2}$ is constant. (d) Ratio $D_\perp/D_\perp^{\rm IE}$; line: 1. Bars are bootstrap standard errors over electrons combined with the sampling error of the reference.}
  \label{fig:nu-D-vs-B}
\end{figure}

The measured rate follows the independent-encounter rate (Fig.~\ref{fig:nu-D-vs-B}a, b, Table~\ref{tab:magnetized}). The reference falls by half from $10^{-4}$ to $5\times10^{-2}\,\mathrm T$ because encounters at impact parameters above the gyroradius hardly deflect the velocity \cite{montgomeryMagneticFieldDependence1974,wareElectronFokkerPlanck1989}. At every field the measured rate is within $10\%$ of the reference and within two standard errors. The cross-field diffusion coefficient, by contrast, exceeds the independent-encounter value by up to a factor of 17 (Section~\ref{subsec:D-scaling}).

\begin{figure}[!htbp]
    \centering
    \PaperFigureWide{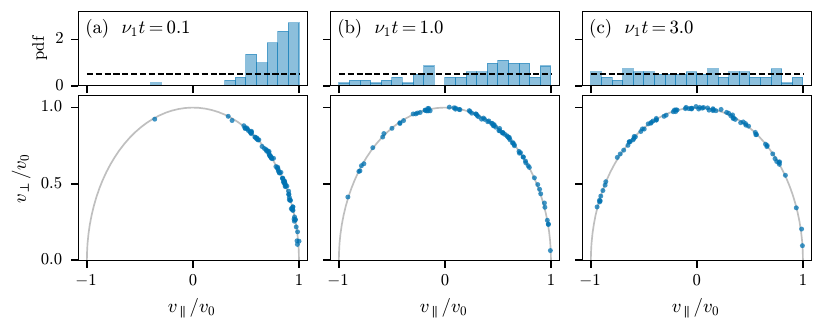}
    \caption{Evolution of the joint velocity distribution in the magnetized simulation at $10^{-2}\,\mathrm T$ (long run), shown in the $(v_\parallel/v_0,\,v_\perp/v_0)$ plane at $\nu_1t=0.1$, 1 and 3 (a--c) for the 82 electrons with initial $\lvert\mu\rvert\ge0.7$; $v_\parallel$ is multiplied by the sign of the initial $\mu$. The top marginals show the corresponding probability density functions of $v_\parallel/v_0$. Gray circle: $\lvert\bv\rvert=v_0$; dashed: isotropic distribution at speed $v_0$. The electrons stay on the circle and spread along it toward the isotropic distribution.}
    \label{fig:vz-vperp-evolution}
\end{figure}

In velocity space the electrons relax toward isotropy at the single rate $\nu$, which at $10^{-2}\,\mathrm T$ is within $10\%$ of the independent-encounter rate. Figure~\ref{fig:vz-vperp-evolution} follows the electrons of the long run at $10^{-2}\,\mathrm T$ (length $40/\nu_1$, Table~\ref{tab:runs}) that start with $\lvert\mu\rvert\ge0.7$, in the $(v_\parallel/v_0,\,v_\perp/v_0)$ plane at three times. They stay on the circle $\lvert\bv\rvert=v_0$, because the ions are static and the magnetic field does no work, and scattering moves them along it toward the isotropic distribution. The mean of $v_\parallel/v_0$, taken with the sign of the initial $\mu$, decays as $\exp(-\nu t)$ with the rate $\nu$ measured in this run, to within $1.4$ standard errors. The departure from independent encounters therefore lies in where the guiding center goes (Section~\ref{subsec:D-scaling}).

\subsubsection{Field-strength scaling and quantitative comparison with theory}\label{subsec:D-scaling}

Independent encounters underestimate the cross-field diffusion coefficient by a factor that grows with the field (Fig.~\ref{fig:nu-D-vs-B}c, d, Table~\ref{tab:magnetized}). The ratio $D_\perp/D_\perp^{\rm IE}$ rises from $1.09\pm0.08$ at $\rho_v=5.6\lambda_D$, where the two agree, to $17\pm2$ at $0.011\lambda_D$ (16 to 25 over the variants of the reference, Appendix~\ref{app:ie}). Measured in the classical unit $D_0$ of each field, $D_\perp$ grows fortyfold from $10^{-4}$ to $5\times10^{-2}\,\mathrm T$, so it falls much more slowly with the field than the classical $B^{-2}$. The reference grows as well, from $1.06$ to $2.78\,D_0$, because more encounters are distant as the gyroradius decreases \cite{oneilNewTheoryTransport1985,dubinTestParticleDiffusion1997}.

\subsection{Origin of the excess}\label{subsec:excess-origin}

Three results show that the excess comes from returns along the same stretch of field line.

\begin{figure}[!htbp]
  \centering
  \PaperFigureFull{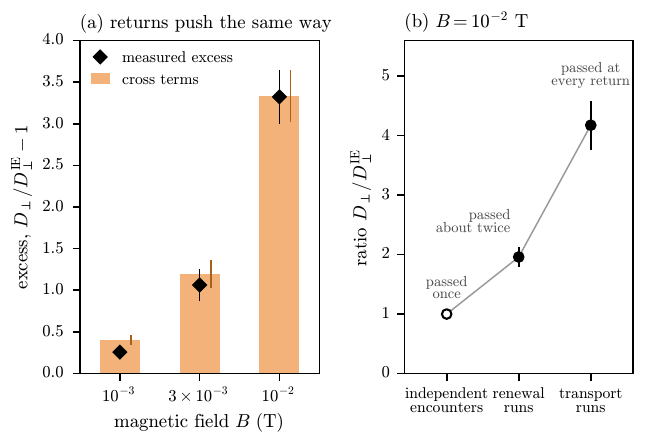}
  \caption{Origin of the excess over independent encounters, at $c=3$. The displacements of different excursions tend to point the same way, and the resulting cross terms are as large as the excess or larger. At $10^{-2}\,\mathrm T$ limiting the repeated passes removes about $70\%$ of it. An excursion is the part of a trajectory between two successive reversals of the parallel velocity, during which the electron moves one way along $\boldsymbol B$. (a) Excess of the transport runs over the independent-encounter reference, $D_\perp/D_\perp^{\rm IE}-1$ with $D_\perp$ from the first line of Eq.~\eqref{eq:excursions} (diamonds), and the cross terms of Eq.~\eqref{eq:excursions} in the same units (bars), at $10^{-3}$, $3\times10^{-3}$ and $10^{-2}\,\mathrm T$. (b) $D_\perp/D_\perp^{\rm IE}$ at $10^{-2}\,\mathrm T$ for the three calculations of Section~\ref{subsec:ie-reference}: the independent-encounter reference (1 by definition, open marker) and, for the same 127 electrons, the renewal runs (ions renewed every $\tau_c=0.03/\nu_1$) and the transport runs (ions fixed; all 256 electrons: $4.4\pm0.4$, Table~\ref{tab:magnetized}). The labels give how often the same ions are passed. Error bars are bootstrap standard errors over electrons.}
  \label{fig:excursions}
\end{figure}

First, the excess lies in the cross terms, not in the single excursions. At each of the three fields of Fig.~\ref{fig:excursions}a the cross terms of Eq.~\eqref{eq:excursions} are positive; they equal the excess $D_\perp/D_\perp^{\rm IE}-1$ within its standard error at $3\times10^{-3}$ and $10^{-2}\,\mathrm T$ and exceed it at $10^{-3}\,\mathrm T$. The simultaneous interactions act while the electron moves one way along the field, so they change the displacement of that excursion but not whether the displacements of different excursions point the same way. They therefore do not cause the excess.

Second, only excursions over the same stretch of field line are correlated. We measure the distance of two excursions along $\boldsymbol B$ as the gap between the stretches of field line they cover. For pairs of excursions that are not consecutive and are less than $\lambda_D$ apart along $\boldsymbol B$, which can pass the same ions, the mean cosine between their displacements is 0.12 to 0.23 over the three fields. For pairs more than $2R_c$ apart, which share no ion, it is below 0.004 in magnitude.

Third, limiting the repeated passes removes most of the excess (Fig.~\ref{fig:excursions}b). For the same 127 electrons at $10^{-2}\,\mathrm T$, $D_\perp/D_\perp^{\rm IE}$ is $4.2\pm0.4$ in the transport runs, where the same ions are passed as often as the orbit returns, and $1.96\pm0.16$ in the renewal runs, where they are passed about twice. The two calculations differ in the repeated passes and in a speed reset at each renewal that keeps the energy (Table~\ref{tab:runs}). Renewing the ion configuration leaves the scattering rate and the single-excursion terms unchanged within their standard errors: the ratio of the rates in the renewal and transport runs is $1.01\pm0.13$, and the single-excursion terms change by $0.02\pm0.06$ in units of $D_\perp^{\rm IE}$. It lowers the cross terms, the tendency of different excursions to move the guiding center the same way. Limiting the repeated passes to about two thus removes about $70\%$ of the excess at $10^{-2}\,\mathrm T$. Section~\ref{sec:discussion} explains why a stronger field gives more passes of the same ions (Fig.~\ref{fig:weak-strong}).

\section{Discussion}
\label{sec:discussion}

\begin{figure}[!htbp]
  \centering
  \PaperSchematic{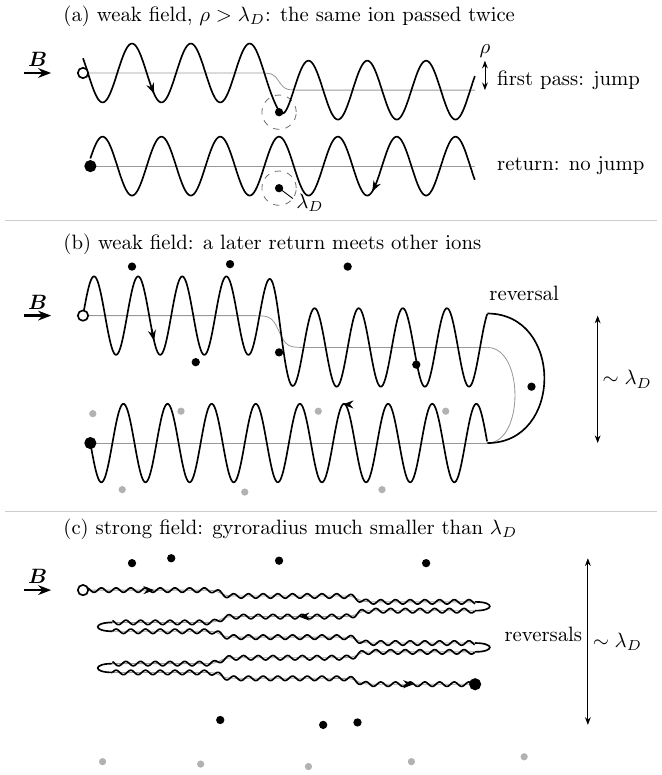}
  \caption{Why the excess grows with the field. In a weak field a return passes the same ion at another gyrophase without repeating the jump (a), and a later return meets other ions (b); in a strong field the electron passes the same ions many times, and each pass shifts its guiding center the same way (c). Side views in a plane that contains $\boldsymbol B$, not to scale. Thick line: helix of one electron; thin gray line: path of its guiding center; dots: ions; open and filled circles: start and end. (a) Weak field, gyroradius larger than $\lambda_D$, the same ion passed twice (drawn one above the other): on the first pass the electron comes within $\lambda_D$ of the ion (dashed circle), and its guiding center jumps; the return, at another gyrophase, passes the ion much farther away. (b) Weak field, a later return: after a scattering time the guiding center has moved by about a gyroradius, more than $\lambda_D$, and the electron passes other ions (gray). (c) Strong field, gyroradius much smaller than $\lambda_D$: the electron reverses several times within about $\lambda_D$ of the same stretch of field line; black dots: ions met on every pass. In (b) and (c) successive passes are drawn apart; only the steps at the ions are guiding-center shifts.}
  \label{fig:weak-strong}
\end{figure}

The excess grows with the field because in a stronger field an electron passes the same ions more often before its guiding center moves away from them (Fig.~\ref{fig:weak-strong}). In a weak field, with the gyroradius comparable to or larger than the Debye length, scattering moves the guiding center by about a gyroradius within a scattering time (Section~\ref{subsec:obs-D}), out of the screening range of the ions it has met (Fig.~\ref{fig:weak-strong}b). An electron that returns within a scattering time meets the same ions mostly in close encounters, and a return at an unrelated gyrophase rarely repeats these (Fig.~\ref{fig:weak-strong}a). Its returns thus add up little: $D_\perp/D_\perp^{\rm IE}$ is $1.09$ at $\rho_v=5.6\lambda_D$ and $1.28$ at $0.56\lambda_D$. In a strong field, with the gyroradius much smaller than the Debye length, the electron is tied to its field line: it reverses many times on the same stretch, passes the same ions each time, and each pass shifts its guiding center the same way (Figs.~\ref{fig:close-distant}b and~\ref{fig:weak-strong}c).

Earlier work proposed this mechanism without measuring it: Polyakov for electrons among static screened charges \cite{polyakovChangeDueScatteringact1986}, and Baalrud and Daligault for the slow decrease of $D_\perp$ with the field in molecular dynamics \cite{baalrudTransportRegimesSpanning2017}. Here the comparison with independent encounters in the same model measures the excess. The orbits locate it in displacements that point the same way on overlapping stretches of field line. Renewing the ion configuration, so that the same ions are passed at most about twice, removes most of it. In a single-species plasma, Dubin found that repeated collisions of the same pair of guiding centers raise the long-range diffusion by a fixed factor of three, independent of the field \cite{dubinTestParticleDiffusion1997}. In Dubin's plasma both particles move along the field. Among static ions the configuration near the field line stays in place, and the excess grows with the field. Molecular dynamics finds long-range correlations along the field from nearly $180^\circ$ collisions at the ends of narrow cylinders around the field lines \cite{vidalExtendedSpaceTime2021}, a picture close to the correlated excursions found here.

For collision operators, the comparison shows which coefficient independent encounters do not reproduce. Without a magnetic field the scattering rate equals the binary-collision rate for the same truncated potential at three interaction radii, consistent with the result of Escande et al.\ that the transport effect of the simultaneous interactions is a sum of independent binary estimates \cite{escandeUniformDerivationCoulomb2015}. Escande et al.\ treat the unmagnetized plasma. With a magnetic field and $\rho_v<\lambda_D$, independent encounters still reproduce the pitch-angle scattering rate to within $10\%$, but they underestimate the cross-field diffusion coefficient by a factor that grows to 17 at $\rho_v=0.011\lambda_D$ (Table~\ref{tab:magnetized}). At $10^{-2}\,\mathrm T$, limiting the repeated passes removes most of the excess (Section~\ref{subsec:excess-origin}). A collision operator for this regime can therefore keep the pitch-angle scattering of independent encounters, with the Coulomb logarithm cut at the gyroradius, but its cross-field diffusion must also include the drift of distant encounters and the repeated passes, to reach $D_\perp$ of Table~\ref{tab:magnetized}, up to $47\,D_0$ at the parameters studied here. Gyrokinetic collision operators must in addition conserve particle number, momentum and energy and satisfy an H theorem \cite{sugamaModernGyrokineticFormulation2017}, as does the linearized multi-species operator of Donnel et al.\ in the full-F code GYSELA \cite{donnelMultispeciesCollisionalOperator2019}. For plasmas in which the ion configuration persists while an electron returns, the pitch-angle scattering rate computed here is a reference value for the deflection term of such operators, and $D_\perp$ is one for the classical flux from finite gyroradii. Gyrokinetic operators that keep finite-gyroradius terms contain this flux \cite{sugamaModernGyrokineticFormulation2017}.

The way the excess depends on how long the ion configuration persists parallels test-particle diffusion in stochastic and turbulent fields, which depends on the correlation time $\tau$ of the field. Diffusion grows with $\tau$ when particles move little during $\tau$, and trapping in the structures of the field changes this when they move far \cite{vladDiffusionIntrinsicTrapping1998,milovanovStochasticDynamicsFractional2001}. In drift-wave turbulence, particles trapped in vortices are carried with them for about the lifetime of the vortices, which enhances their diffusion \cite{naulinDispersionIdealParticles1999}. The static ions form a field with an infinite correlation time, and renewing the ion configuration makes this time finite. Likewise, the ions near the field line of an electron shift its guiding center the same way for as long as their configuration persists. It persists when the ions are heavy and cold, as in magnetized ultracold neutral plasmas \cite{gormanMagneticConfinementUltracold2021}.

\section{Conclusion}
\label{sec:conclusion}

We developed a framework for electron--ion collisional transport that integrates test-electron Newton--Lorentz trajectories in the direct Debye-screened field of static ions, and compared the resulting transport with an independent-encounter reference computed in the same model. The use of a time-symmetric pusher, midpoint field evaluation, and adaptive time stepping makes it possible to extract long-time transport observables with encounters resolved down to head-on passages, and a persistent ion background lets a returning electron meet the same ions.

Without a magnetic field, the pitch-angle scattering rate agrees with the binary-collision rate for the same truncated potential at three interaction radii, to $0.2\%\pm0.7\%$ on average (Section~\ref{subsec:unmagnetized_nu}). For gyroradii from 5.6 to $0.011\lambda_D$, the scattering rate stays within $10\%$ of the independent-encounter reference at every field, whereas the cross-field diffusion coefficient exceeds its reference by a factor that grows from 1.3 at $\rho_v=0.56\lambda_D$ to 17 at $0.011\lambda_D$ (Section~\ref{subsec:D-scaling}). Splitting each trajectory where its velocity along the field reverses shows that the excess comes from repeated passes of the same ions, and that the guiding-center displacements on these passes point the same way (Section~\ref{subsec:excess-origin}). Renewing the ion configuration every 0.03 scattering times limits these passes to about two and lowers $D_\perp/D_\perp^{\rm IE}$ at $\rho_v=0.056\lambda_D$ ($10^{-2}\,\mathrm T$) from 4.2 to 1.96 for the same electrons.

The results describe electron--ion transport wherever the ion configuration persists while an electron returns along the field, as for static scatterers and approximately for heavy, cold ions. The present calculation complements reduced collisional transport theory by testing the independence of successive encounters directly against screened Coulomb dynamics, and the same framework can include mobile ions and electron--electron collisions. For collision models, independent encounters give the pitch-angle scattering rate, but not the cross-field diffusion coefficient once the gyroradius is below the Debye length, because an electron that turns back along the field passes the same ions again.

\section*{Acknowledgments}
This work is supported by the National Magnetic Confinement Fusion Energy Research and Development Program of China (No. 2024YFE03020004) and the Geo-Algorithmic Plasma Simulator (GAPS) Project.

\section*{Data availability}
The data that support the findings of this study are openly available in Zenodo at \url{https://doi.org/10.5281/zenodo.21301653} (version 2.0.0). The release includes processed numerical data underlying the reported figures and results, simulation metadata, and scripts to reproduce the reported figures and results. It also includes a Python implementation of Algorithm~\ref{ttl:alg}, which reproduces trajectories of the production code to round-off. Simulations were performed using the Accurate Particle Tracer (APT) code and research-specific extensions; the APT source code is not included because the authors are not authorized to redistribute it. The complete simulation outputs are available from the corresponding author on reasonable request.

\section*{CRediT authorship contribution statement}
\textbf{Keheng Zhu:} Conceptualization, Methodology, Software, Investigation,
Formal analysis, Validation, Data curation, Visualization,
Writing -- original draft, Writing -- review \& editing.
\textbf{Jian Liu:} Conceptualization, Methodology, Supervision, Algorithms,
Project administration, Funding acquisition, Resources, Writing -- review \& editing.
\textbf{Chaozhou Mou:} Validation, Writing -- review \& editing.
\textbf{Senran Lin:} Validation, Writing -- review \& editing.
\textbf{Wei Zhang:} Validation, Writing -- review \& editing.

\section*{Declaration of competing interest}
The authors declare that they have no known competing financial interests or personal relationships that could have appeared to influence the work reported in this paper.

\section*{Declaration of generative AI and AI-assisted technologies
in the manuscript preparation process}
During the preparation of this work, the authors used ChatGPT
(OpenAI) for language editing and improving readability.
After using this tool, the authors reviewed and edited the content
as needed and take full responsibility for the content of the article.

\appendix
\makeatletter
\renewcommand{\thesubsection}{\@Alph\c@section.\arabic{subsection}}
\@addtoreset{table}{section}\@addtoreset{figure}{section}
\makeatother
\section{Implementation of the Step}\label{ttl:app:proofs}

The step is a second-order, time-symmetric time-transformed leapfrog \cite{MikkolaAarseth2002} whose drifts follow the gyration exactly \cite{Patacchini2009}, with the auxiliary variable $W$ of Mikkola and Aarseth setting the physical duration of the drifts. The step function $g$ of Eq.~\eqref{ttl:eq:g} combines the single-ion terms in a fourth-power sum \cite[Appendix~A]{Hands2019}, \cite[Section~2.3]{Boekholt2023}, with the taper of Steinbach and Brooks \cite{SteinbachBrooks1994}. Algorithm~\ref{ttl:alg} gives the integration of one electron.

\begin{algorithm}[!htbp]
\caption{Integration of one electron through the output times $t_1<t_2<\dots<t_K$. $\Hel(\tau)$ is the exact motion in the uniform magnetic field alone over the physical time $\tau$. The force $\bF=q_e\boldsymbol E$ of Eq.~\eqref{eq:superposed-field}, $g$ and $\nabla g$ are sums over the listed ions within $R_c$ of the point given, evaluated in one loop. The ion background is fixed and generated cell by cell, so it needs no storage: cubic cells of side $R_c/2$ each hold a Poisson-distributed number of ions with mean $n(R_c/2)^3$, at uniform random positions generated from a seed derived from the cell index, a global seed and the index of the electron.}
\label{ttl:alg}
\begin{algorithmic}[1]
\Require position $\bx$ and velocity $\bv$ at $t=0$; ion background; $\Delta\tilde t$, $g_\infty$, $\ell$
\State $t\gets0$,\quad $W\gets g(\bx)$,\quad $k\gets1$
\While{$k\le K$}
  \State $(\bx_s,\bv_s,t_s)\gets(\bx,\bv,t)$ \Comment{state at the start of the step}
  \State neighbor list: the ions within $R_c+\ell$ of $\bx_s$
  \State $\tau_1\gets\Delta\tilde t/(2W)$;\quad $(\bx,\bv)\gets\Hel(\tau_1)(\bx,\bv)$;\quad $t\gets t+\tau_1$ \Comment{half drift}
  \State $\tau_K\gets\Delta\tilde t/g(\bx)$;\quad $\bv^+\gets\bv+\tau_K\bF(\bx)/m_e$ \Comment{kick}
  \State $W\gets W+\tau_K\,\tfrac12(\bv+\bv^+)\cdot\nabla g(\bx)$;\quad $\bv\gets\bv^+$
  \State $\tau_2\gets\Delta\tilde t/(2W)$;\quad $(\bx,\bv)\gets\Hel(\tau_2)(\bx,\bv)$;\quad $t\gets t+\tau_2$ \Comment{half drift}
  \State stop the electron if $W\le0$ or $\lvert\bv\rvert$ exceeds $10^{-2}$ of the speed of light
  \While{$k\le K$ and $t_k\le t$} \Comment{output times passed by this step}
    \State $\Delta t_{\rm out}\gets t_k-t_s$;\quad $(\bx',\bv')\gets\Hel(\Delta t_{\rm out}/2)(\bx_s,\bv_s)$ \Comment{on a copy}
    \State $\bv'\gets\bv'+\Delta t_{\rm out}\,\bF(\bx')/m_e$;\quad $(\bx',\bv')\gets\Hel(\Delta t_{\rm out}/2)(\bx',\bv')$
    \State record $(\bx',\bv')$ as the state at $t_k$;\quad $k\gets k+1$
  \EndWhile
\EndWhile
\end{algorithmic}
\end{algorithm}

\section{Estimators and Independent-Encounter Reference}\label{app:reference}

\subsection{Estimators}\label{app:estimators}

Table~\ref{tab:estimators} lists the settings of the estimators of Section~\ref{subsec:transport-benchmarks}.

\begin{table}[!htbp]
\centering
\caption{Settings of the estimators; $\nu_r=\nu_0$ at $\boldsymbol B=0$ and $\nu^{\rm IE}$ in a field.}
\label{tab:estimators}
\footnotesize
\setlength{\tabcolsep}{4pt}
\begin{tabular}{@{}>{\raggedright\arraybackslash}p{2.4cm}>{\raggedright\arraybackslash}p{10.6cm}@{}}
\toprule
$\nu$, Eq.~\eqref{eq:nu-fit} & least-squares fit of $\ln C$ of the unit vector $\hat{\mathbf u}$ over $0.05\le\nu_r\tau\le1$, time origins $s\ge0.05/\nu_r$ \\\addlinespace
$D_\perp$, Eq.~\eqref{eq:Dperp} & slope from the time interval $\tau=1/\nu_1$, after which the initial pitch angle is forgotten, to the largest interval, about $6/\nu_1$, with the same time origins \\\addlinespace
Uncertainties & bootstrap over electrons, 2000 resamples, $\nu$ and $D_\perp$ in the same resample \\\addlinespace
Stopped electrons & electrons that reach $10^{-2}$ of the speed of light are stopped (Algorithm~\ref{ttl:alg}): one at $c=0.5$, $B=0$, kept up to $\nu_0t=0.86$; one at $10^{-2}\,\mathrm T$, replaced by an additional electron; one in the long run \\
\bottomrule
\end{tabular}
\end{table}

\subsection{Coulomb Logarithm of the Truncated Potential}\label{app:coulomb-log}

On a straight path at impact parameter $b$, an ion acts only along the chord of the sphere of radius $R_c$ about it.
The transverse impulse is the bare-Coulomb value $2m_ev_0b_0/b$ multiplied by
\begin{equation}
f(x;c)=\int_0^{\beta_m}(\cos\beta+x)\,e^{-x/\cos\beta}\,d\beta,\qquad \beta_m=\arctan\frac{(c^2-x^2)^{1/2}}{x},
\label{eq:impulse-factor}
\end{equation}
for $x=b/\lambda_D<c$, and $f=0$ for $x\ge c$. The factor is $f=1$ for the bare Coulomb field and $f=xK_1(x)$ for
$c\to\infty$, where $K_1$ is a modified Bessel function. The constant of Eq.~\eqref{eq:lnL-eff} is
\begin{equation}
K(c)=\lim_{x_0\to0}\Bigl[\ln x_0+\int_{x_0}^{c}f(x;c)^2\,\frac{dx}{x}\Bigr],
\label{eq:K-const}
\end{equation}
which tends to $\ln2-\gamma-\tfrac12=-0.384$ as $c\to\infty$, where $\gamma$ is Euler's constant~\cite{liboff1959,escandeUniformDerivationCoulomb2015}. For
$c=0.5$, 1 and 3, $K=-1.273$, $-0.754$ and $-0.401$.

\subsection{Independent-Encounter Reference}\label{app:ie}

Slow passes need a cutoff. As $\mu\to0$ the drift of a pass grows as $1/\lvert\mu\rvert$ and the sum over passes,
Eq.~\eqref{eq:nuIE}, does not converge, whereas among the other ions scattering ends such a pass early. Following
Dubin \cite{dubinTestParticleDiffusion1997}, we drop the drift of passes with $\mu<\mu_c(b)=(\nu^{\rm IE}b/v_0)^{1/3}$,
which would not end before scattering changes $\mu$ by about $\mu$. The shift of the guiding center by the kicks
within $\rho_v/2$ of the ion, the close-encounter jump, is always kept.

The cutoff has two parameters, the threshold $\mu_c$ and the radius $\rho_v/2$. We varied them: $\mu_c$ multiplied by
2 or $\tfrac12$, replaced by a uniform threshold from 0.02 to 0.15, or computed with the measured $\nu$, and the radius
doubled to $\rho_v$. These variants change $D_\perp^{\rm IE}$ by at most $32\%$, at $5\times10^{-2}\,\mathrm T$, much
less than the excess: there $D_\perp/D_\perp^{\rm IE}$ is 16 to 25 over the variants. Uncertainties come from a
jackknife over 64 batches of passes.

\FloatBarrier%
\bibliographystyle{elsarticle-num}
\bibliography{Plasma-Collision,APT,TTL,Magnetized}

\end{document}